\newif\ifsingle
\singletrue 

\ifsingle
\documentclass[11pt,draftclsnofoot, onecolumn]{IEEEtran}		
\else		
\documentclass[10pt,final, twocolumn]{IEEEtran}
\usepackage[a4paper, total={7.3in, 10.4in}]{geometry}
\fi

\usepackage{times}
\usepackage{amsmath,dsfont}
\usepackage{amssymb,amsthm}
\usepackage{epsfig,verbatim}
\usepackage{subfigure}
\usepackage{setspace}
\usepackage{color}
\usepackage{cite}
\usepackage{epstopdf}
\usepackage{enumitem}
\usepackage{graphics}
\usepackage{accents}
\usepackage{acronym}
\usepackage[bookmarks,colorlinks]{hyperref}
\usepackage{booktabs}
\usepackage{mathtools}

\definecolor{NewColor2}{rgb}{0,0,0}
\definecolor{NewColor}{rgb}{0,0,0}

\newcommand{\myVec}[1]{{\boldsymbol{#1}}}
\newcommand{\myMat}[1]{{\boldsymbol{#1}}}
\newcommand{\mySet}[1]{\mathcal{#1}}

\newcommand{\E}{\mathds{E}}		 			
\newcommand{\myI}{{\myMat{I}}}			 		
\newcommand{\myX}{{\myVec{x}}}			 		
\newcommand{\myTh}{{\myVec{s}}}			 		
\newcommand{\myThTil}{\tilde{\myTh}}			 		

\newcommand{\myThTilEnt}{\tilde{s}}			 		
\newcommand{\Rate}{R}

\newcommand{\myPS}{a}			 		

\newcommand{\DynRange}{\gamma}
\newcommand{\DynInt}[1][\lenZ]{\Delta_{#1}}

\newcommand{\Pdf}[1]{f_{ #1}}
\newcommand{\lenX}{n}			 			
\newcommand{\lenS}{k}			 			
\newcommand{\lenZ}{p}			 			
\newcommand{\lenSset}{\mySet{K}}			 			
\newcommand{\Quan}[2]{Q_{#1}^{#2}}
\newcommand{\myC}{{\myVec{c}}}			 		
\newcommand{\CovMat}[1]{\myMat{\Sigma}_{#1}}			
\newcommand{\eig}[1]{\lambda_{#1}}			
\newcommand{\LmmseMat}{\myMat{\Gamma}}

\newcommand{\Vecdim}[1]{} 

\newcommand{\TilM}[1][\lenZ]{\tilde{M}_{#1}}
\newcommand{\myW}{w}			 		

\newcommand{\MyKappa}[1][\lenZ]{\kappa_{#1}}
\newcommand{\Qnoise}{\myVec{e}}
\newcommand{\Wlevel}{\zeta}
\newcommand{\SigX}{\sigma_{\myX,{\rm max}}^2}
\newcommand{\SigS}{\sigma_{\myThTil,{\rm max}}^2}

\newcommand{\myA}{\myMat{A}}				
\newcommand{\myB}{\myMat{B}}

\newcommand{\opt}{^{\rm o}}			
\newcommand{\Hanalog}{h_{\rm a}}
\newcommand{\Hdigital}{h_{\rm d}}
\newcommand{\myEta}{\eta}

\newcommand{\Ny}{n_{\rm x}}
\newcommand{\Ky}{m_{\rm x}}
\newcommand{\Ns}{n_{\rm s}}
\newcommand{\Ks}{m_{\rm s}}
\newcommand{\myXi}{\myX}
\newcommand{\myZ}{\myVec{z}}

\acrodef{rv}[RV]{random variable}
\acrodef{mse}[MSE]{mean-squared error}
\acrodef{bs}[BS]{base station}
\acrodef{dft}[DFT]{discrete Fourier transform} 
\acrodef{adc}[ADC]{analog-to-digital convertor}
\acrodef{dt}[DT]{discrete-time}
\acrodef{ct}[CT]{continuous-time}
\acrodef{awgn}[AWGN]{additive white Gaussian noise}
\acrodef{wss}[WSS]{wide-sense stationary}
\acrodef{mmse}[MMSE]{minimum \ac{mse}}
\acrodef{lmmse}[LMMSE]{linear \ac{mmse}}
\acrodef{mimo}[MIMO]{multiple-input multiple-output}
\acrodef{map}[MAP]{maximum a-posteriori probability} 
\acrodef{isi}[ISI]{intersymbol interference}
\acrodef{snr}[SNR]{signal-to-noise ratio}
\acrodef{svd}[SVD]{singular value decomposition}
\acrodef{pc}[PC]{proper-complex}
\acrodef{pdf}[PDF]{probability density function}
\acrodef{doa}[DOA]{direction of arrival}

\newtheorem{definition}{Definition}
\newtheorem{theorem}{Theorem}
\newtheorem{corollary}{Corollary}
\newtheorem{proposition}{Proposition}
\newtheorem{lemma}{Lemma}

\ifsingle
\newcommand{\includefig}[1]{\includegraphics[width = 0.75\columnwidth]{#1} 	\vspace{-0.4cm}}
\else
\newcommand{\includefig}[1]{\includegraphics[width = \columnwidth]{#1} 	\vspace{-0.6cm}}
\fi 

\long\def\symbolfootnote[#1]#2{\begingroup\def\thefootnote{\fnsymbol{footnote}}\footnote[#1]{#2}\endgroup}

 \IEEEoverridecommandlockouts
\title{Hardware-Limited Task-Based Quantization
}

\vspace{-0.2cm}

\author{
	\IEEEauthorblockN{ Nir Shlezinger, Yonina C. Eldar, and Miguel R. D. Rodrigues\\
	}
	\thanks{Parts of this work were accepted for presentation in the 2019 IEEE International Workshop on Signal Processing Advances in Wireless Communications (SPAWC), Cannes, France.
	}
	\thanks{This project has received funding from the European Union’s Horizon 2020 research and innovation program under grant No. 646804-ERC-COG-BNYQ, from the Israel Science Foundation under grant No. 0100101, and from the Royal Society International Exchange scheme IE 160348.
	}
	\thanks{N. Shlezinger  and Y. C. Eldar are with the faculty of Math and CS, Weizmann Institute of Science,  Israel (e-mail: nirshlezinger1@gmail.com; yonina@weizmann.ac.il). 	
	}
	\thanks{M. R. D. Rodrigues  is with the department of EE,  University College, London, UK (e-mail: m.rodrigues@ucl.ac.uk).  	
	}

	\vspace{-0.7cm}
	
}

\vspace{-0.5cm}

\begin{document}

\maketitle
\pagestyle{plain}
\thispagestyle{plain}
\begin{abstract}
	Quantization plays a critical role in  digital signal processing systems. 
	Quantizers are typically designed to obtain an accurate digital representation of the input signal,  operating independently of the system task, and are commonly implemented using serial scalar analog-to-digital converters (ADCs). 
	In this work, we study hardware-limited task-based quantization, where a system utilizing a serial scalar ADC is designed to provide a suitable representation in order to allow the recovery of a parameter vector underlying the input signal.
%
	We propose hardware-limited task-based quantization systems for a fixed and finite quantization resolution, and characterize their achievable distortion. 	 
	We then apply the analysis to the practical setups of channel estimation and eigen-spectrum recovery from quantized measurements.
Our results illustrate that properly designed hardware-limited systems can approach the optimal performance achievable with vector quantizers, and that by taking the underlying task  into account, the quantization error can be made negligible with a relatively small number of bits.
\end{abstract}

\vspace{-0.6cm}
\section{Introduction}
\vspace{-0.1cm}
Quantization refers to the representation of a continuous-amplitude signal using a finite dictionary, or equivalently, a finite number of bits \cite{Gray:98}. 
Quantizers are implemented in digital signal processing systems using \acp{adc}, which typically operate in a serial scalar manner due to hardware-limitations. In such systems, each incoming continuous-amplitude sample is represented in digital form using the same mechanism \cite{Eldar:15}.
The quantized representation is commonly selected to accurately match the original signal, in the sense of minimizing some distortion measure, such that the signal can be recovered with minimal error from the quantized measurements  \cite[Ch. 10]{Cover:06}, \cite{Berger:98}.

 Quantization design is typically performed regardless of the system task. However, in many signal processing applications, the goal is not to recover the actual signal, but to capture certain parameters, such as an underlying model or unknown channel, 
from the quantized signal \cite{Rodrigues:17}.  We refer to systems where one wishes to extract some information from the quantized signal, rather than recovering the signal itself, as {\em task-based quantization}, and to such systems operating with serial scalar \acp{adc}  as {\em hardware-limited task-based quantization} systems.  

\label{txt:IntHL}
Hardware-limited quantization with low resolution is the focus of growing interest over recent years due to the increasing complexity and bitrate demands of modern signal processing and communications systems.  
Common tasks considered with low resolution hardware-limited quantization include \ac{mimo} communications \cite{Rini:17,Choi:18,Choi:17, Mo:17,Li:17,Choi:16,Zhang:16}, channel estimation    \cite{Zeitler:12,Mo:18,Dabeer:10,Stien:18,Li:17,Choi:16,Sung:18}, subspace estimation \cite{Chi:17}, time difference of arrival estimation  \cite{Corey:17}, and \ac{doa} estimation  \cite{Yu:16,Liu:17}.  
In these works it is assumed that quantization is carried out {\em separately from the system task}, typically using {\em fixed uniform} low-precision quantizers, e.g., one-bit quantization of a scalar value is implemented using the ${\rm sign}$ function \cite{Rini:17}. 
\textcolor{NewColor}{
These quantized measurements are then processed in the digital domain using different inference algorithms, such as linear estimators \cite{Li:17, Dabeer:10}, maximum-likelihood estimation \cite{Stien:18}, and approximate message passing based algorithms \cite{Mo:18,Zhang:16,Sung:18}. }
 As these works focus only on the digital processing, they do not provide guidelines to designing quantization systems with a small and finite number of bits by acknowledging the task of the system.

When hardware-limitations are not present,  task-based quantization systems can take advantage of joint vector quantization, which is known to be superior to serial scalar quantization \cite[Ch. 22.2]{Polyanskiy:15}. 
Previous works on task-based quantization without hardware limitations are divided according to whether the parameter vector is modeled as a random vector, namely, a Bayesian setup, or as a deterministic unknown parameter. 
When the signal parameter is random, task-based quantization can be viewed as an indirect lossy source coding problem\footnote{Direct lossy source coding typically refers to the standard quantization setup where the task of the system is to recover the quantized signal, while indirect source coding refers to task-based quantization \cite{Kostina:16}.} \cite[Sec. V-G]{Gray:98}. 
For this setup with a stationary source that is related to the observation vector via a stationary memoryless channel, Witsenhausen proved in  \cite{Witsenhausen:80} that the rate-distortion function, namely, the minimal number of bits required to obtain a given representation accuracy determined by the distortion measure, is asymptotically equivalent to the rate-distortion function for representing the observed signal -- instead of the signal parameter -- with a surrogate distortion measure. Under \ac{mse} distortion, Wolf and Ziv proved in \cite{Wolf:70} that this equivalence also holds for finite signal size, and the work \cite{Rodrigues:17} provided guidelines to the optimal joint quantization and estimation scheme. Recently, Kostina and Verdu characterized nonasymptotic bounds on the rate-distortion functions for indirect as well as direct lossy source coding with arbitrary distortion measures  \cite{Kostina:16, Kostina:12}, by considering single-shot quantization, and specialized the bounds for i.i.d. signals with separable distortion. 
The indirect source coding framework was also used to study conversion of continuous-time signals into quantized discrete-time signals in \cite{Kipnis:18, Kipnis:18a}.
The focus in the works \cite{Witsenhausen:80,Wolf:70,Kostina:12, Kostina:16,Kipnis:18,Kipnis:18a} is on the 
{\em optimal} tradeoff between quantization rate and achievable distortion. Consequently, their results cannot be applied to quantify the achievable performance of practical hardware-limited systems utilizing serial scalar \acp{adc}. 

For a signal parameter modeled as deterministic and unknown, \cite{Gupta:03} studied detection from quantized observations, i.e., recovering a scalar binary parameter, while \cite{Varshney:08} treated detection from quantized prior probabilities. Quantization for the recovery of a scalar parameter taking values on a discrete finite set was studied in  \cite{Dogahe:11}. 
The design of quantizers for the recovery of a vector parameter taking values on a continuous set was considered in \cite{Bianchi:13}, which proposed an adaptive algorithm for tuning the quantizer.  In all the works above, i.e., \cite{Gupta:03,Bianchi:13,Dogahe:11,Varshney:08}, as well as in \cite{Rodrigues:17}, the analysis assumes vector quantizers with high resolution, where the number of bits used for representing the quantized signal can be made arbitrarily large. They do not consider practical systems that utilize serial scalar quantizers with a fixed and finite number of bits.

\vspace{-0.25cm}
\subsection {Main Contributions}
\vspace{-0.1cm}
\label{txt:Contributions}
In this work we study quantization systems which are hardware-limited to utilize practical serial scalar \acp{adc} operating with a fixed  number of bits, for the task of acquiring a random parameter vector taking values on a continuous set. Our approach is to account for the task in design of the quantization system, thus mitigating the effect of the structure imposed by hardware limitations. This hardware-limited task-based framework fits a broad range of signal processing and communications systems. In particular, a task may be any data processing or inference objective, and various constraints encountered in practice, such as sampling rate limitations, delay restrictions, and imposed structures, can be represented as hardware limitations. Consequently, although our derivation here considers quantization for the task of estimating a desired parameter vector while restricted to using scalar \acp{adc}, the underlying approach, i.e., to overcome hardware limitations by properly accounting for a task in the system design, is applicable to a much broader family of systems.

We first consider the case where the observations and the desired vector are related such that the \ac{mmse} estimate is a linear function of the observations. Such relationships are commonly encountered in channel estimation and signal recovery problems,  e.g., \cite{Rodrigues:17,Choi:17, Mo:17, Li:17, Choi:16, Zeitler:12, Zhang:16,Mo:18, Dabeer:10,Stien:18}. 
We focus on \textcolor{NewColor}{structured} systems implementing uniform quantization with linear processing, allowing analog combining prior to digital processing. 
 This approach was previously studied in the context of \ac{mimo} communications  \cite{Rini:17,Stein:17, AlKhateeb:14, Cuba:17}.
For this setup, we derive the \textcolor{NewColor} {hardware-limited task-based quantization system which minimizes the \ac{mse}}, and characterize the achievable distortion. 
The proposed system accounts for the task by reducing the number of quantized samples via an appropriate linear transformation to be not larger than the size of the desired signal. It then  rotates the quantized samples to have identical variance. 
Quantization is performed based on a waterfilling-type expression, accounting for the serial operation and the limited dynamic range of practical \acp{adc}.

In addition, we  characterize the minimal achievable distortion of two suboptimal approaches: We first discuss systems in which processing is carried out only in the digital domain, as is the structure considered in the majority of the literature of tasks performed with low resolution quantization, e.g., \cite{Zeitler:12,Mo:18,Corey:17,Dabeer:10,Stien:18,Yu:16, Chi:17}. Then, we study systems which quantize the \ac{mmse} estimate, an approach which is known to be optimal when using vector quantizers \cite{Wolf:70}, and was also proposed for compressed sensing with quantized measurements \cite{Kipnis:17}.
Surprisingly, we show that, unlike when vector quantizers are employed, in the presence of serial scalar \acp{adc}, quantizing the \ac{mmse} estimate is generally suboptimal. We provide a necessary and sufficient condition for this approach to coincide with the optimal design. 

Next, we extend the proposed system to scenarios where  the observations and the desired vector are related via an arbitrary stochastic model. In particular, we identify the main design guidelines associated with the case where the \ac{mmse} estimate is a linear function of the observations, and discuss how they may be applied for arbitrary models. Then, we explicitly show how these guidelines can be used to construct a  hardware-limited task-based quantization system for scenarios in which the desired vector can be recovered from the empirical covariance of the observations, as in \cite{Rodrigues:17,Chi:17,Corey:17, Yu:16,Liu:17}. 

Finally, we apply our results to two practical setups: Channel estimation from quantized measurements \cite{Zeitler:12,Mo:18,Dabeer:10,Stien:18,Li:17,Choi:16} and eigen-spectrum estimation from quantized measurements\cite{Rodrigues:17}. 
We demonstrate that, by properly accounting for the presence of serial scalar \acp{adc}, practical  hardware-limited systems operating with a relatively small number of bits approach the optimal performance, achievable with vector quantizers, in practical and relevant scenarios. 
Furthermore, we show that hardware-limited quantizers designed accounting for the task of the system can substantially outperform task-ignorant systems utilizing vector quantizers. This gain is mainly achieved by applying task-based linear analog processing, in addition to the digital processing.

\vspace{-0.25cm}
\subsection {Organization and Notations}
\vspace{-0.1cm}
The rest of this paper is organized as follows: 
Section~\ref{sec:Preliminaries} briefly reviews some preliminaries in quantization theory, and formulates the hardware-limited task-based quantization setup. 
Section~\ref{sec:Pre_Task} discusses task-based quantization with vector quantizers.   
Section~\ref{sec:Hardware} studies hardware-limited task-based quantization when the \ac{mmse} estimate is linear, 
and Section~\ref{sec:Asymptotic} extends the proposed design to arbitrary setups.
Section~\ref{sec:Simulations} presents the application of the results  in a numerical study. Section \ref{sec:Conclusions} provides some concluding remarks.
Detailed proofs of the results are given in the appendix.

Throughout the paper, we use boldface lower-case letters for vectors, e.g., ${\myVec{x}}$;
the $i$th element of ${\myVec{x}}$ is written as $({\myVec{x}})_i$. 
Matrices are denoted with boldface upper-case letters,  e.g., 
$\myMat{M}$, and  $(\myMat{M})_{i,j}$   is its $(i,j)$th element. 
Sets are denoted with calligraphic letters, e.g., $\mySet{X}$, and $\mySet{X}^n$ is the $n$th order Cartesian power of $\mySet{X}$. 
Transpose,  Euclidean norm, trace, stochastic expectation,  sign, and mutual information are written as  $(\cdot)^T$,  $\left\|\cdot\right\|$, ${\rm {Tr}}\left(\cdot\right)$, $\E\{ \cdot \}$, $ {\rm sign}\left(\cdot\right)$,  and $I\left( \cdot ~ ; \cdot \right)$, respectively, 
and $\mySet{R}$ is the set of real numbers. We use $a^+$ to denote $\max(a,0)$, and $\myI_{n}$ is the $n \times n$ identity matrix. 
All logarithms are taken to basis 2.

\vspace{-0.2cm}
\section{Preliminaries and Problem Statement}
\label{sec:Preliminaries}
\vspace{-0.1cm}
%
\subsection{Preliminaries in Quantization Theory}
\label{subsec:Pre_Quant}
\begin{figure}
	\centering
	\ifsingle
	\includegraphics[width = 0.6\columnwidth]{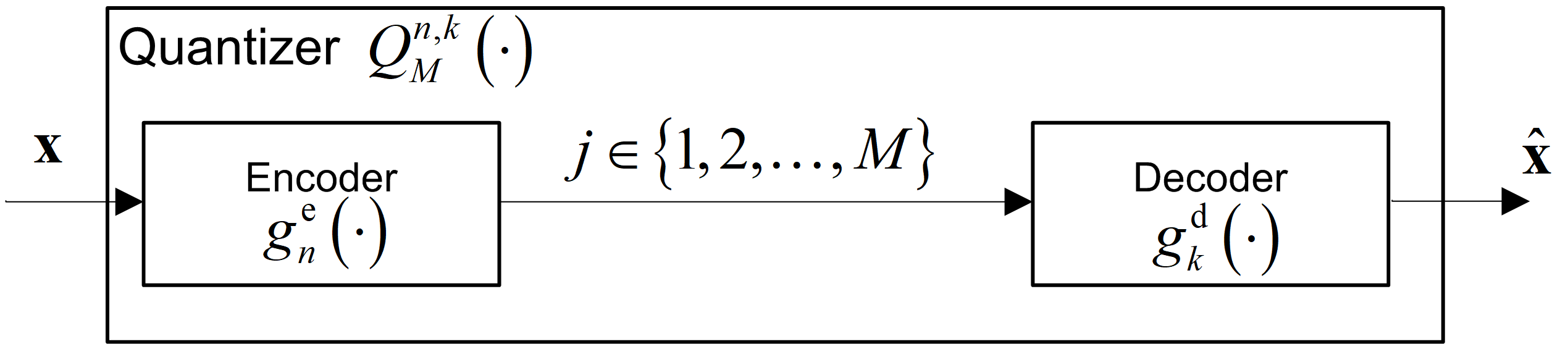}
	\else
	\includefig{fig/Classical_RD3.png}
	\fi
	\caption{Quantizer illustration.}
	\vspace{-0.3cm}
	\label{fig:StandardIllust}
\end{figure}

\vspace{-0.1cm}
To formulate the hardware-limited task-based quantization problem, we first review standard quantization notations, after which we discuss task-based quantization.
To that aim, we recall the definition of a quantizer:
\begin{definition}[Quantizer]
	\label{def:Quantizer}
	A quantizer $\Quan{M}{n,k}\left(\cdot \right)$ with $\log M$ bits, input size $n$, input alphabet $\mySet{X}$, output size $k$, and output alphabet $\hat{\mySet{X}}$, consists of: 
	{\em 1)} An  encoding function $g_n^{\rm e}: \mySet{X}^n \mapsto \{1,2,\ldots,M\} \triangleq \mySet{M}$ which maps the input into a discrete index $i \in \mySet{M}$.
	{\em 2)} A decoding function  $g_k^{\rm d}: \mySet{M} \mapsto \hat{\mySet{X}}^k$ which maps each index $i \in \mySet{M}$ into a codeword $\myVec{q}_i \in  \hat{\mySet{X}}^k$. 
\end{definition}
We write the output of the quantizer with input $\myX\Vecdim{n} \in \mySet{X}^n$ as $\hat{\myX}\Vecdim{k} = g_k^{\rm d}\left( g_n^{\rm e}\left( \myX\Vecdim{n}\right) \right) \triangleq \Quan{M}{n,k}\left( \myX\Vecdim{n}\right)$. 
 {\em Scalar quantizers} operate on a scalar input, i.e., $n=1$ and $\mySet{X}$ is a scalar space, while {\em vector quantizers} have a multivariate input. 
The set of codewords $ \{\myVec{q}_i\}_{i=1}^M$ is referred to as the {\em quantization codebook}. 
When the input size and  output size are equal, namely, $n=k$, we write $\Quan{M}{n}\left(\cdot \right) \triangleq \Quan{M}{n,n}\left(\cdot \right)$. An illustration is given in Figure \ref{fig:StandardIllust}. 

\smallskip
\subsubsection{Standard Quantization}
In the standard quantization problem, a $\Quan{M}{n}\left(\cdot \right)$ quantizer is designed to minimize some distortion measure  $d_n:\mySet{X}^n\times\hat{\mySet{X}}^n \mapsto \mySet{R}^+$  between its input and its output. 
The performance of a quantizer is therefore characterized using two measures: The quantization rate, defined as $\Rate \triangleq \frac{1}{n}\log M$, and the expected distortion $\E\{d_n\left(\myX\Vecdim{n}, \hat{\myX}\Vecdim{n} \right)\}$. For a fixed input size $n$ and codebook size $M$, the optimal quantizer is thus given by  
\begin{equation}
\label{eqn:OptQuantizer}
\Quan{M}{n, {\rm opt}}\left(\cdot \right) = \mathop{\arg \min}\limits_{\Quan{M}{n}\left(\cdot \right)} \E \left\{d_n\left(\myX, \Quan{M}{n}\left( {\myX} \right)\right)   \right\}.
\end{equation}

Characterizing the optimal quantizer via \eqref{eqn:OptQuantizer} and the optimal tradeoff between distortion and quantization rate is in general a very difficult task. Consequently, optimal quantizers are typically studied assuming either high quantization rate, i.e., $\Rate \rightarrow \infty$, see, e.g., \cite{Li:99}, or asymptotically large input size, namely, $n \rightarrow \infty$, typically with i.i.d. inputs\footnote{Rate-distortion theory can also be used for non i.i.d. signals, see, e.g., \cite[Ch. 5]{Han:03}. However, the simple classical expression, as given by the distortion-rate function in Def. \ref{def:DistRateFunction}, requires the observed signal to have i.i.d. entries.}, via rate-distortion theory \cite[Ch. 10]{Cover:06}. 
For example, when the quantizer input consists of i.i.d. \acp{rv} with probability measure $\Pdf{x}$, and the distortion measure can be written as $d_n\left(\myX, \hat{\myX} \right) = \frac{1}{n} \sum\limits_{i=1}^n d\left(\left( \myX \right) _i , \left( \hat{\myX} \right) _i\right)$ for some  $d:\mySet{X}\times\mySet{X} \mapsto \mySet{R}^+$, then the optimal distortion in the limit $\lenX \rightarrow \infty$ for a given rate $\Rate$ is given by the distortion-rate function:
 \begin{definition}[Distortion-rate function]
 	\label{def:DistRateFunction}
 	The distortion-rate function for input $x \in \mySet{X}$ with respect to the distortion measure $d:\mySet{X}\times\mySet{X} \mapsto \mySet{R}^+$ is defined as
 	\vspace{-0.1cm}
 	\begin{equation}
 	\label{eqn:DistRateFunction}
 	D_{x}\left( R \right) = \mathop {\min }\limits_{{\Pdf{\hat{x}|x}}:I\left( \hat{x};x \right) \le R} \E\left\{ {d\left( \hat{x},x \right)} \right\}.
 	\vspace{-0.1cm}
 	\end{equation}
 \end{definition}
 The conditional distribution which obtains the minima in \eqref{eqn:DistRateFunction}, $\Pdf{\hat{\myX}|\myX}^{\rm opt}$, is referred to as the {\em optimal distortion-rate distribution}, and the marginal distribution $\Pdf{\hat{\myX}} =  \int \Pdf{\hat{\myX}|\myX}^{\rm opt}\Pdf{\myX}$ is referred to henceforth as the {\em optimal marginal distortion-rate distribution}.

Comparing high quantization rate analysis for scalar quantizers and rate-distortion theory for vector quantizers demonstrates the sub-optimality of serial scalar quantization. For example, for  large $\Rate$, even for i.i.d. inputs, vector quantization outperforms serial scalar quantization, with a distortion gap of $4.35$ dB for Gaussian inputs with the \ac{mse} distortion \cite[Ch. 23.2]{Polyanskiy:15}.

\smallskip
\subsubsection{Task-Based Quantization} 
\begin{figure}
	\centering
	\includefig{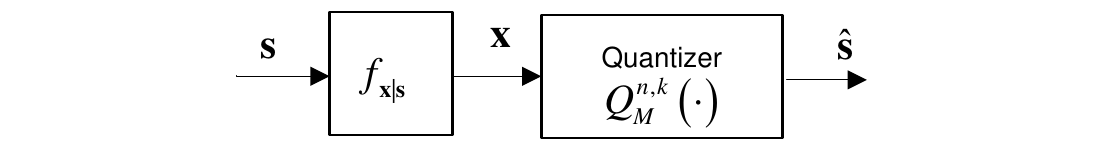}
	\caption{Task-based quantization illustration.}
	\vspace{-0.4cm}
	\label{fig:Illustration2}
\end{figure}

In task-based quantization the design objective of the quantizer is some task 
other than minimizing the distortion between its input and output.  
\label{txt:FiniteVar}
\textcolor{NewColor}{
In the following, we focus on the generic task of acquiring a zero-mean random vector  $\myTh \in \mySet{R}^\lenS$ from a measured zero-mean random vector $\myX  \in \mySet{R}^\lenX$, where the entries of $\myX$ and  $\myTh$ have finite variance,  and $\lenX \ge \lenS > 0$.}
This formulation accommodates a broad range of tasks, including channel estimation, covariance estimation, and source localization. A natural distortion measure for such setups is the \ac{mse}, which we consider throughout the paper. An illustration of a task-based quantization system is depicted in Figure \ref{fig:Illustration2}.


\vspace{-0.2cm} 
\subsection{Problem Formulation}
\label{subsec:ProblemFormulation}
\begin{figure}
	\centering
	\ifsingle
	\includegraphics{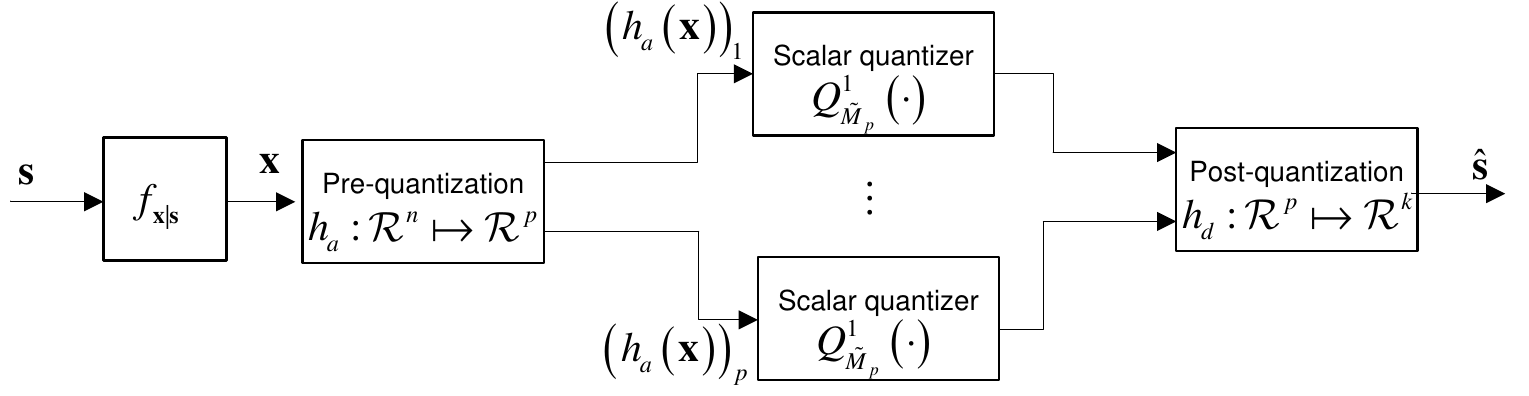}
	\else
	\includefig{GenericStructure2.pdf}
	\fi
	\caption{Hardware-limited task-based quantizer.}
	\vspace{-0.6cm}
	\label{fig:HardLimIllust}
\end{figure}

\vspace{-0.1cm} 
In this work we study task-based quantization with hardware limitations.
As discussed in the introduction, practical digital signal processing systems typically obtain a digital representation of physical analog signals using serial scalar \acp{adc}. 
We refer to task-based quantization with serial scalar \acp{adc} as {\em hardware-limited task-based quantization}. 
Since in such systems, each continuous-amplitude sample is converted into a discrete representation using a single quantization rule, this operation can be modeled using {\em identical scalar quantizers}.  Consequently, the system we consider is modeled using the setup depicted in Fig. \ref{fig:HardLimIllust}.
The observed signal $\myX \in \mySet{R}^\lenX$ is projected into $\mySet{R}^\lenZ$, $\lenZ \le \lenX$, using some mapping $\Hanalog(\cdot)$, which represents the pre-quantization processing carried out in the analog domain. 
Since general mappings may be difficult to implement in analog, we henceforth restrict $\Hanalog(\cdot)$ to be a linear function, namely, we only allow {\em linear analog combining}, as in, e.g., \cite{Rini:17,Stein:17}.  In this case, $\Hanalog(\myX) = \myA \myX$ for some $\myA \in \mySet{R}^{\lenZ \times \lenX}$. 

%
Each entry of $\Hanalog \big(\myX\Vecdim{\lenX} \big)$ is quantized using the same scalar quantizer with resolution $\TilM \triangleq \lfloor M^{1/\lenZ} \rfloor$, denoted $\Quan{\TilM}{1}(\cdot)$. The overall number of quantization levels is thus $\big(\TilM\big)^\lenZ \le M$. 
We note that  $M$, which represents the memory requirement of the system, is also directly related to the \ac{adc} power consumption. However, for the same overall number of quantization levels $M$, different selections of $\lenZ$ may result in different power consumptions, depending on the physical implementation of the \ac{adc}, see, e.g., \cite{Cuba:17}. In the following we keep the value of $M$ fixed and finite, i.e., the memory requirement, which is independent of the specific implementation of the \ac{adc}, is the same for all considered systems.

The representation of  $\myTh\Vecdim{\lenS}$, denoted $\hat{\myTh}\Vecdim{\lenS}$, is obtained as the output of some post-quantization mapping $\Hdigital:\mySet{R}^\lenZ \mapsto \mySet{R}^\lenS$, applied to the output of the identical scalar quantizers. The mapping $\Hdigital(\cdot)$ represents the joint-processing carried out in the digital domain. 
The quantized representation $\hat{\myTh}\Vecdim{\lenS}$ can be written as
\begin{equation}
\label{eqn:QuantRep}
\hat{\myTh}\Vecdim{\lenS} = \Hdigital\left(\Quan{\TilM}{1}\big( \left(\Hanalog\left(\myX \right)  \right)_1\big), \ldots, \Quan{\TilM}{1}\big( \left(\Hanalog\left(\myX \right)  \right)_\lenZ\big) \right).
\end{equation}
%

The novelty of the model in Fig. \ref{fig:HardLimIllust}, compared to previous works on quantization for specific tasks with serial scalar \acp{adc}, e.g., \cite{Choi:17, Mo:17,Li:17,Choi:16, Zeitler:12,Mo:18,Dabeer:10,Stien:18,Chi:17, Corey:17, Yu:16,Liu:17,Zhang:16,Sung:18}, is in the introduction of the additional linear processing carried out in the analog domain, represented by the mapping $\Hanalog(\cdot)$. 
The concept of using analog combining prior to digital processing was previously studied in the context of \ac{mimo} communications in \cite{Rini:17,Choi:18, Stein:17,AlKhateeb:14,Cuba:17}.
The motivation for introducing  $\Hanalog(\cdot)$ is to reduce the dimensionality of the input to the \ac{adc}, thus facilitating a more accurate quantization without increasing the overall number of bits, $\log M$. As shown in the following sections, by properly designing $\Hanalog(\cdot)$, this approach can substantially improve the performance of task-based quantizers operating with serial scalar \acp{adc}.

Our analysis of hardware-limited task-based quantization, focusing on the \ac{mse} distortion,  
consists of three parts: 
\begin{enumerate}
	\item As a preliminary step, in Section \ref{sec:Pre_Task}, we discuss non hardware-limited task-based quantization systems, namely, systems implementing task-based quantization using optimal vector quantizers instead of serial scalar \acp{adc}. The purpose of this analysis is to serve as a basis for comparing the performance of hardware-limited task-based quantizers to vector quantizers.
	\item Next, in Section \ref{sec:Hardware}, we focus on the case where $\myX$ and $\myTh$ are related such that the \ac{mmse} estimate of $\myTh$ from $\myX$ is a linear function of $\myX$. Such relationships arise in various channel estimation and signal recovery setups, e.g., \cite{Rodrigues:17, Choi:17, Mo:17, Li:17, Choi:16, Zeitler:12, Mo:18,Zhang:16, Dabeer:10,Stien:18}.For this setting, we propose a hardware-limited quantization system design, and characterize its achievable distortion. 	We also characterize the minimal achievable distortion when no pre-quantization processing is carried out, as well as when the analog combiner is designed to recover the \ac{mmse} estimate. 
	\item Then, in Section \ref{sec:Asymptotic}, we use the characterization of $\Hanalog(\cdot)$ and $\Hdigital(\cdot)$ given in Section \ref{sec:Hardware} for linear models, to provide guidelines for designing $\Hanalog(\cdot)$ and $\Hdigital(\cdot)$ under arbitrary relationships between $\myX$ and $\myTh$. We suggest a concrete design for cases in which $\myTh$ can be estimated from the second-order statistical moments of $\myX$, as in \cite{Rodrigues:17,Chi:17,Corey:17, Yu:16,Liu:17}.
\end{enumerate}
Our analysis shows that, unlike when vector quantizers are applied, the optimal strategy for systems utilizing serial scalar \acp{adc} is not to quantize the \ac{mmse} estimate. Instead, the input to the \ac{adc} is rotated to account for the identical quantization rule of serial scalar \acp{adc}, and includes a waterfilling-type expression to account for the limited dynamic range.  Furthermore, our numerical comparison presented in Section~\ref{sec:Simulations} demonstrates that the proposed system, which uses simple hardware, can approach the performance of the optimal  vector quantizer.

\vspace{-0.25cm}
\section{Task-Based  Vector Quantization}
\label{sec:Pre_Task}
\begin{figure}
	\centering
	\includefig{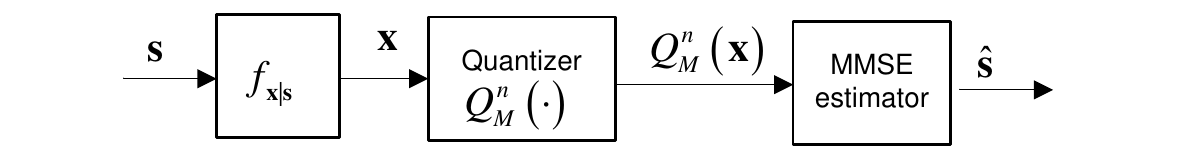}
	\caption{Task-ignorant quantizer.}
	\vspace{-0.6cm}
	\label{fig:Illustration4}
\end{figure}

\vspace{-0.1cm}
As a preliminary step towards our study of hardware-limited task-based quantizers, we consider task-based quantization which utilizes vector quantizers without hardware limitations.
%
%
We focus on two approaches for task-based quantization: 
In the first, referred to as {\em optimal task-based quantization}, the quantizer $\Quan{n,k}{M}(\cdot)$ in Figure \ref{fig:Illustration2}  is designed to recover the desired vector $\myTh\Vecdim{\lenS}$.  
In the second strategy, described in Figure \ref{fig:Illustration4} and referred to as {\em task-ignorant quantization}, the quantizer is designed  to recover the observed vector $\myX\Vecdim{\lenX}$ separately from the task, and  $\myTh\Vecdim{\lenS}$ is estimated from the quantized representation. 
The optimal task-based quantizer obtains the minimal achievable distortion for a given quantization rate, while the task-ignorant quantizer  represents the best system one can construct  when the quantizer is designed separately from the task. 
The approaches we discuss below are based on joint vector quantization, and thus cannot be implemented using practical serial scalar \acp{adc}.

\subsubsection{Optimal Task-Based Quantizer} 
\label{subsub:Optimal}
For the \ac{mse} distortion, the optimal quantizer is   constructed by first obtaining the \ac{mmse} estimate of $\myTh$ from $\myX$,  ${\myThTil\Vecdim{\lenS}} \!=\! \E\left\{ \left. {{\myTh \Vecdim{\lenS}}} \right|\myX\Vecdim{\lenX} \right\}$, and then quantizing the estimate  \cite{Wolf:70}. This leads to a minimal distortion given by 
\ifsingle
\vspace{-0.1cm}
\begin{align} 
&\mathop {\min }\limits_{\Quan{M}{\lenX,\lenS}\left(  \cdot  \right)} \E\left\{ \left\| {{\myTh \Vecdim{\lenS}}-{\Quan{M}{\lenX,\lenS}}\left( \myX\Vecdim{\lenX} \right)} \right\|^2 \right\} 
= \E\left\{ \left\| {{\myTh \Vecdim{\lenS}}-{\myThTil\Vecdim{\lenS}}} \right\|^2 \right\} 
+ \mathop {\min }\limits_{\Quan{M}{\lenS}\left(  \cdot  \right)} \E\left\{ \left\| {{\myThTil \Vecdim{\lenS}}-\Quan{M}{\lenS}\left( {{\myThTil\Vecdim{\lenS}}} \right)} \right\|^2 \right\}.
\vspace{-0.1cm}
\label{eqn:VecQuan1}
\end{align} 
\else
\vspace{-0.1cm}
\begin{align} 
&\mathop {\min }\limits_{\Quan{M}{\lenX,\lenS}\left(  \cdot  \right)} \E\left\{ \left\| {{\myTh \Vecdim{\lenS}}-{\Quan{M}{\lenX,\lenS}}\left( \myX\Vecdim{\lenX} \right)} \right\|^2 \right\} 
= \E\left\{ \left\| {{\myTh \Vecdim{\lenS}}-{\myThTil\Vecdim{\lenS}}} \right\|^2 \right\} 
\notag \\
&\qquad \qquad \qquad \qquad  
+ \mathop {\min }\limits_{\Quan{M}{\lenS}\left(  \cdot  \right)} \E\left\{ \left\| {{\myThTil \Vecdim{\lenS}}-\Quan{M}{\lenS}\left( {{\myThTil\Vecdim{\lenS}}} \right)} \right\|^2 \right\}.
\vspace{-0.1cm}
\label{eqn:VecQuan1}
\end{align} 
\fi 
It follows from \eqref{eqn:VecQuan1} that the minimal distortion is the sum of the minimal estimation error  of $\myTh\Vecdim{\lenS}$ from  $\myX\Vecdim{\lenX}$, and the minimal distortion in quantizing the \ac{mmse} estimate ${\myThTil \Vecdim{\lenS}}$.  The latter can be obtained explicitly under a high quantization rate assumption, i.e., $\frac{1}{\lenS} \log M \rightarrow \infty$, using fine quantization analysis, as was done in \cite{Rodrigues:17}, or alternatively, when $\myThTil$ has i.i.d. entries and $\lenS$ tends to infinity, using rate-distortion theory.
For finite $\lenS$, $\lenX$ and $M$, the minimal distortion in quantizing the \ac{mmse} estimate may be bounded  as stated in the following proposition:
\begin{proposition}
	\label{pro:JointOpt1} 
	For any random vector $\tilde{\myC}\Vecdim{\lenS} \in \mySet{R}^\lenS$ with probability measure $\Pdf{\tilde{\myC}\Vecdim{\lenS}}$ independent of  $\myThTil\Vecdim{\lenS}$, the minimal \ac{mse} in quantizing the \ac{mmse} estimate $\myThTil\Vecdim{\lenS}$ using a $\Quan{M}{\lenS}$ quantizer satisfies
\ifsingle
	\vspace{-0.1cm}
	\begin{align}
	&D_{\myThTil\Vecdim{\lenS}}(\log M) \le \mathop {\min }\limits_{\Quan{M}{\lenS}\left(  \cdot  \right)} \E\left\{ \left\| {{\myThTil \Vecdim{\lenS}}\!-\!\Quan{M}{\lenS}\left( {{\myThTil\Vecdim{\lenS}}} \right)} \right\|^2 \right\} 
	\le \E \left\{ \int\limits_0^\infty  \left[ \Pr \left( \left. \left\| \tilde{\myC}\Vecdim{\lenS} \!-\! \myThTil\Vecdim{\lenS} \right\|^2 > t \right|\myThTil\Vecdim{\lenS} \right) \right]^M dt  \right\}.
	\vspace{-0.1cm}
	\label{eqn:JointOpt1}
	\end{align}
\else
	\vspace{-0.1cm}
	\begin{align}
	&D_{\myThTil\Vecdim{\lenS}}(\log M) \le \mathop {\min }\limits_{\Quan{M}{\lenS}\left(  \cdot  \right)} \E\left\{ \left\| {{\myThTil \Vecdim{\lenS}}\!-\!\Quan{M}{\lenS}\left( {{\myThTil\Vecdim{\lenS}}} \right)} \right\|^2 \right\} 
	 		\notag \\
	 		&\qquad \qquad 
	\le \E \left\{ \int\limits_0^\infty  \left[ \Pr \left( \left. \left\| \tilde{\myC}\Vecdim{\lenS} \!-\! \myThTil\Vecdim{\lenS} \right\|^2 > t \right|\myThTil\Vecdim{\lenS} \right) \right]^M dt  \right\}.
	\vspace{-0.1cm}
	\label{eqn:JointOpt1}
	\end{align}
\fi 
\end{proposition}

{\em Proof:}
See Appendix \ref{app:Proof1}.
%

\smallskip 
The bounds in \eqref{eqn:JointOpt1} are used in the sequel for comparing the performance of hardware-limited task-based quantization to the optimal performance achievable using vector quantizers. 
\textcolor{NewColor2}{These bounds are required since evaluating the minimal distortion of a vector quantizer, i.e., the middle term in \eqref{eqn:JointOpt1}, is a challenging task for finite signal size and quantization resolution.}
The upper bound in \eqref{eqn:JointOpt1} is the exact performance of random coding, which is known to provide a relatively tight bound  for fixed blocklengths \cite{Kostina:12}, and to asymptotically achieve the distortion-rate curve \cite[Ch. 23.2]{Polyanskiy:15}.  
A reasonable assignment for  $\Pdf{\tilde{\myC}\Vecdim{\lenS}}$ in \eqref{eqn:JointOpt1}  is the optimal marginal distortion-rate distribution; with this distribution the distortion of quantizers with i.i.d. random codewords coincides with the distortion-rate function for sources generating asymptotically large number of  i.i.d. realizations of ${\myThTil \Vecdim{\lenS}}$  \cite[Ch. 24.2]{Polyanskiy:15}. 
\textcolor{NewColor2}{Alternative upper bounds, which may be tighter and simpler to compute compared to the one used in \eqref{eqn:JointOpt1} yet are not analytically expressible, can be obtained by numerically evaluating the distortion of iterative and data-driven vector quantizer designs, such as the extension of Lloyd's algorithm to multivariate inputs \cite{Linde:80}.}
In general, the distortion-rate function and the optimal marginal  distribution can be obtained using iterative algorithms, e.g.,  the Blahut-Arimoto algorithm \cite[Ch. 10.8]{Cover:06} and its extensions to continuous-valued \acp{rv} \cite{Dauwels:05}.

\subsubsection{Task-Ignorant Quantizer} 
\label{subsub:Ignorant}
When the quantizer operates independently of the task, the desired vector $\myTh$  must be estimated directly from the quantized observations. For the optimal quantizer and estimator for this setup,  $\Quan{M}{\lenX}(\cdot)$  minimizes the \ac{mse} between its output and  $\myX\Vecdim{\lenX}$, and  $\myTh\Vecdim{\lenS}$ is estimated from the output of the quantizer using the \ac{mmse} estimator. 
From the orthogonality principle,  the resulting \ac{mse} in estimating $\myTh\Vecdim{\lenS}$ is 
\ifsingle 
\begin{align} 
\E\left\{ \left\| {\myTh \Vecdim{\lenS}}-\E\left\{\myTh\Vecdim{\lenS} \big|  {\Quan{M}{\lenX}}\left( \myX\Vecdim{\lenX} \right) \right\}\right\|^2 \right\} 
&= \E\left\{ \left\| {{\myTh \Vecdim{\lenS}}-{\myThTil\Vecdim{\lenS}}} \right\|^2 \right\} 
+   \E\left\{ \left\| {\myThTil \Vecdim{\lenS}}\! - \!
\E\left\{\myTh\Vecdim{\lenS} \big|  {\Quan{M}{\lenX}}\left( \myX\Vecdim{\lenX} \right) \right\} \right\|^2 \right\} \notag \\
&\stackrel{(a)}{=} \E\left\{ \left\| {{\myTh \Vecdim{\lenS}}\! - \!{\myThTil\Vecdim{\lenS}}} \right\|^2 \right\} +  \E\left\{ \left\| {\myThTil \Vecdim{\lenS}}\! - \!
\E\left\{\myThTil\Vecdim{\lenS} \big|  {\Quan{M}{\lenX}}\left( \myX\Vecdim{\lenX} \right) \right\} \right\|^2 \right\},
\vspace{-0.1cm}
\label{eqn:Unconst1}
\end{align} 
\else
\begin{align} 
&\E\left\{ \left\| {\myTh \Vecdim{\lenS}}-\E\left\{\myTh\Vecdim{\lenS} \big|  {\Quan{M}{\lenX}}\left( \myX\Vecdim{\lenX} \right) \right\}\right\|^2 \right\} \notag \\
&\qquad
= \E\left\{ \left\| {{\myTh \Vecdim{\lenS}}-{\myThTil\Vecdim{\lenS}}} \right\|^2 \right\} 
+   \E\left\{ \left\| {\myThTil \Vecdim{\lenS}}\! - \!
\E\left\{\myTh\Vecdim{\lenS} \big|  {\Quan{M}{\lenX}}\left( \myX\Vecdim{\lenX} \right) \right\} \right\|^2 \right\} \notag \\
&\qquad\stackrel{(a)}{=} \E\left\{ \left\| {{\myTh \Vecdim{\lenS}}\! - \!{\myThTil\Vecdim{\lenS}}} \right\|^2 \right\} +  \E\left\{ \left\| {\myThTil \Vecdim{\lenS}}\! - \!
\E\left\{\myThTil\Vecdim{\lenS} \big|  {\Quan{M}{\lenX}}\left( \myX\Vecdim{\lenX} \right) \right\} \right\|^2 \right\},
\vspace{-0.1cm}
\label{eqn:Unconst1}
\end{align} 
\fi 
where  $(a)$ follows since $\myTh\Vecdim{\lenS} \mapsto \myX\Vecdim{\lenX} \mapsto  {\Quan{M}{\lenX}}\left( \myX\Vecdim{\lenX} \right) $ form a Markov chain, thus, by \cite[Prop. 4]{Rioul:10}, 
$\E\left\{\myTh\Vecdim{\lenS} \big|  {\Quan{M}{\lenX}}\left( \myX\Vecdim{\lenX} \right) \right\} 
= \E\left\{\myThTil\Vecdim{\lenS} \big|  {\Quan{M}{\lenX}}\left( \myX\Vecdim{\lenX} \right) \right\}$.
The relation in \eqref{eqn:Unconst1} shows that the distortion of the task-ignorant quantizer is given by the sum of the estimation error of the \ac{mmse} estimate  $\myThTil\Vecdim{\lenS}$ and the estimation error of the {\em \ac{mmse} estimate of $\myThTil\Vecdim{\lenS}$ from the quantizer output ${\Quan{M}{\lenX}}\left( \myX\Vecdim{\lenX} \right) $}. 
The main difference between \eqref{eqn:Unconst1} and the optimal estimation error in \eqref{eqn:VecQuan1} is that in \eqref{eqn:Unconst1} the quantizer is fixed, while in \eqref{eqn:VecQuan1} it can be set to minimize the estimation error. 

In order to compute \eqref{eqn:Unconst1}, the distribution of $\myThTil\Vecdim{\lenS}$ given ${\Quan{M}{\lenX}}\left( \myX\Vecdim{\lenX} \right) $ is required, which may be difficult to characterize. 
One scenario in which this requirement can be relaxed is when    $\myThTil\Vecdim{\lenS}$ is a linear function of $\myX\Vecdim{\lenX}$, i.e., $\myThTil = \LmmseMat\myX\Vecdim{\lenX}$ for some $\LmmseMat \in \mySet{R}^{\lenS \times \lenX}$. 
To formulate the resulting distortion, 	let $\CovMat{\myX\Vecdim{\lenX} }$ and $\CovMat{{\Quan{M}{\lenX}}\left( \myX\Vecdim{\lenX} \right)}$ be the covariance matrices of $\myX\Vecdim{\lenX}$ and of ${\Quan{M}{\lenX}}\left( \myX\Vecdim{\lenX} \right) $, respectively. 
The \ac{mse} in this case is stated in the following proposition.
\begin{proposition}
	\label{pro:TaskIgnLinear}
	When $\myThTil\Vecdim{\lenS}  = \LmmseMat\myX\Vecdim{\lenX}$, $\myX\Vecdim{\lenX}$ is zero-mean, and ${\Quan{M}{\lenX}}\left( \myX\Vecdim{\lenX} \right)$ is the optimal quantizer of $\myX\Vecdim{\lenX}$, then
	\begin{align}
	\hspace{-0.2cm}
	\E\!\left\{ \!\left\| {\myThTil \Vecdim{\lenS}} \!-\!
	\E\!\left\{\myThTil\Vecdim{\lenS} \big|  {\Quan{M}{\lenX}}\!\left( \myX\Vecdim{\lenX} \right)\! \right\} \right\|^2 \right\}
	\!=\! {\rm Tr} \left(\LmmseMat^T \LmmseMat \left(\CovMat{\myX\Vecdim{\lenX} } \!-\! \CovMat{{\Quan{M}{\lenX}}\left( \myX\Vecdim{\lenX} \right)} \!\right) \right). 	
	\label{eqn:TaskIgnLinear}
	\end{align}
\end{proposition}

\begin{IEEEproof}
	When $\myThTil\Vecdim{\lenS}$  is the linear \ac{mmse} estimator, 
	the second summand in \eqref{eqn:Unconst1} can be written as $ \E\!\left\{\! \left\| \LmmseMat \!\left( \myX\Vecdim{\lenX}\!-\!\E\left\{\myX\Vecdim{\lenX}  \big| {\Quan{M}{\lenX}}\!\left( \myX\Vecdim{\lenX} \right)\right\} \right) \!\right\|^2 \right\}$. Therefore,
\ifsingle  
	\begin{align}
	\E\!\left\{ \!\left\| {\myThTil \Vecdim{\lenS}} \!-\!
	\E\left\{\myThTil\Vecdim{\lenS} \big|  {\Quan{M}{\lenX}}\!\left( \myX\Vecdim{\lenX} \right) \right\} \right\|^2 \right\} 
	&\stackrel{(a)}{=}\E\left\{ \left\| \LmmseMat \left( \myX\Vecdim{\lenX}\!-\!{\Quan{M}{\lenX}}\left( \myX\Vecdim{\lenX} \right)\right)  \right\|^2 \right\} \notag \\
	&= {\rm Tr} \left(\LmmseMat^T \LmmseMat \E\left\{\left( \myX\Vecdim{\lenX}\!-\!{\Quan{M}{\lenX}}\left( \myX\Vecdim{\lenX} \right)\right)\left( \myX\Vecdim{\lenX}\!-\!{\Quan{M}{\lenX}}\left( \myX\Vecdim{\lenX} \right)\right)^T    \right\}\right) \notag \\
	&\stackrel{(b)}{=} {\rm Tr} \left(\LmmseMat^T \LmmseMat \left(\CovMat{\myX\Vecdim{\lenX} } \!-\! \CovMat{{\Quan{M}{\lenX}}\left( \myX\Vecdim{\lenX} \right)} \!\right) \right),
	\end{align}
\else
	\begin{align}
	&\E\!\left\{ \!\left\| {\myThTil \Vecdim{\lenS}} \!-\!
	\E\left\{\myThTil\Vecdim{\lenS} \big|  {\Quan{M}{\lenX}}\!\left( \myX\Vecdim{\lenX} \right) \right\} \right\|^2 \right\} 
	\stackrel{(a)}{=}\E\left\{ \left\| \LmmseMat \left( \myX\Vecdim{\lenX}\!-\!{\Quan{M}{\lenX}}\left( \myX\Vecdim{\lenX} \right)\right)  \right\|^2 \right\} \notag \\
	&\qquad= {\rm Tr} \left(\LmmseMat^T \LmmseMat \E\left\{\left( \myX\Vecdim{\lenX}\!-\!{\Quan{M}{\lenX}}\left( \myX\Vecdim{\lenX} \right)\right)\left( \myX\Vecdim{\lenX}\!-\!{\Quan{M}{\lenX}}\left( \myX\Vecdim{\lenX} \right)\right)^T    \right\}\right) \notag \\
	&\qquad\stackrel{(b)}{=} {\rm Tr} \left(\LmmseMat^T \LmmseMat \left(\CovMat{\myX\Vecdim{\lenX} } \!-\! \CovMat{{\Quan{M}{\lenX}}\left( \myX\Vecdim{\lenX} \right)} \!\right) \right),
	\end{align}
\fi 
	where $(a)$ follows since ${\Quan{M}{\lenX}}\left( \myX\Vecdim{\lenX} \right)$ is the optimal quantizer of $\myX\Vecdim{\lenX}$ in the \ac{mse} sense, hence  ${\Quan{M}{\lenX}}\left( \myX\Vecdim{\lenX} \right) = \E\big\{\myX\Vecdim{\lenX}  \big| {\Quan{M}{\lenX}}\left( \myX\Vecdim{\lenX} \right)\big\}$; and $(b)$ is a result of the fact that the optimal quantizer is uncorrelated with the quantization error \cite[Sec. III]{Gray:98}. 
\end{IEEEproof}

\smallskip
Proposition \ref{pro:TaskIgnLinear} suggests that, when $\myThTil\Vecdim{\lenS}$ is a linear function of $\myX\Vecdim{\lenX}$,  the distortion can be evaluated using only the covariance matrix of the  task-ignorant quantizer ${\Quan{M}{\lenX}}\left( \myX\Vecdim{\lenX} \right)$. Nonetheless, the covariance of the quantizer which minimizes the distortion with respect to $\myX\Vecdim{\lenX}$ is typically difficult to compute for finite $M$. 
Since  $I(\myX\Vecdim{\lenX} ; \Quan{M}{\lenX}(\myX)) \!\le\! \log M$ \cite[Thm. 23.2]{Polyanskiy:15}, a possible approach to {\em approximate}  the distortion is to evaluate Proposition \ref{pro:TaskIgnLinear} with the covariance matrix of the  output distribution which obtains the  distortion-rate function
$D_{\myX\Vecdim{\lenX}}\left( \log M\right)$, instead of $\CovMat{{\Quan{M}{\lenX}}\left( \myX\Vecdim{\lenX} \right)}$. This replacement provides a reliable characterization of the performance of random codes distributed via the optimal marginal distortion-rate distribution for large  $M$. In the numerical study in Section \ref{sec:Simulations} we illustrate that \eqref{eqn:TaskIgnLinear} approaches the performance of the optimal quantizer designed to recover $\myX\Vecdim{\lenX}$.

\vspace{-0.25cm}%
\section{Hardware-Limited Task-Based Quantization Systems Design}
\label{sec:Hardware}
\vspace{-0.15cm}
\subsection{Model Assumptions}
\label{subsec:SysModel}
\vspace{-0.1cm}
We now study the design of hardware-limited task-based quantization systems illustrated in Fig. \ref{fig:HardLimIllust}. 
\color{NewColor} 
As stated in the problem formulation, we consider the case where $\lenX$, $\lenS$ and $\log M$ are fixed and finite, namely, we do not assume high quantization rate or arbitrarily large inputs. In such cases, explicitly characterizing the optimal quantization system and the minimal achievable distortion is a very difficult task, just as characterizing the minimal achievable distortion in lossy source coding with fixed blocklengths is difficult \cite{Kostina:12, Kostina:16}. 
Consequently, in the following section we focus on scenarios in which the stochastic relationship between the vector of interest $\myTh$ and the observation vector $\myX$ are such that the \ac{mmse} estimate of $\myTh$ from $\myX$, $\myThTil = \E \{\myTh | \myX\}$, is a linear function of $\myX$.  	
Additionally,  we restrict the digital mapping $\Hdigital(\cdot)$ to be linear, namely, $\Hdigital(\myVec{u}) = \myB \myVec{u}$,  $\myB \in \mySet{R}^{\lenS \times \lenZ}$. Since  the \ac{mmse} estimate is linear here,  this constraint is not expected to have a notable effect on the overall performance, especially when the error due to quantization is small. 	

By focusing on these setups, we are able to explicitly derive the achievable distortion and to characterize the system which achieves  minimal distortion. This derivation reveals some non-trivial insights. For example, we show that the optimal approach when using vector quantizers, namely, to quantize the \ac{mmse} estimate, is no longer optimal in the presence of serial scalar \acp{adc}. Furthermore, as detailed in Section \ref{sec:Asymptotic}, this analysis provides general guidelines for designing hardware-limited task-based quantization systems, which can be used for any relationship between $\myTh$ and $\myX$.
 
\color{black}
 


To design a system which operates with simple scalar uniform quantizers, we carry out our analysis assuming dithered quantization \cite{Gray:93}.  Using dithered quantizers results in some favorable properties of the quantized signal, elaborated on in the sequel, 
which facilitate the analysis. These properties are also {\em approximately} satisfied without dithering for many input distributions \cite{Widrow:96}. Therefore, by considering dithered quantization, we are able to rigorously derive the optimal system, where in practice the resulting system can approach the optimal performance using standard uniform quantizers without dithering. 

\begin{figure}
	\centering
	\scalebox{0.85}
	{\includefig{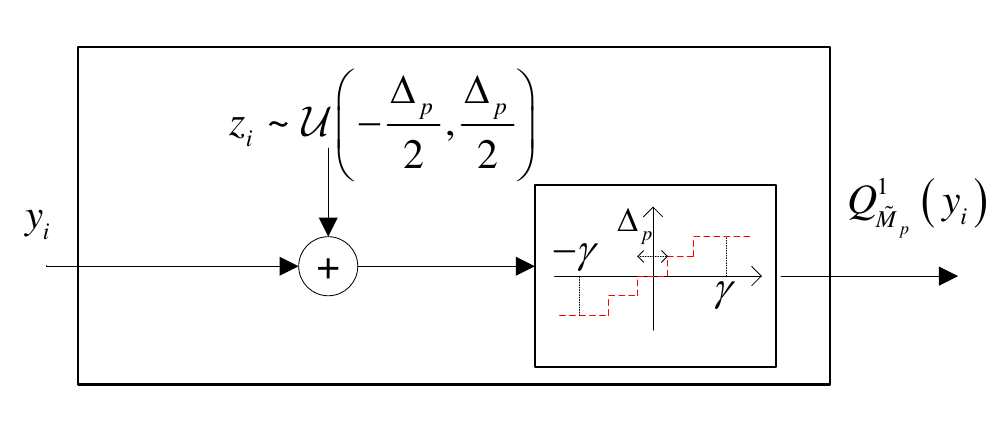}}
	\vspace{-0.4cm}
	\caption{Dithered uniform quantization illustration.}
	\label{fig:DithQuan}		
	\vspace{-0.6cm}
\end{figure}

More specifically, we assume the identical scalar quantizers $\Quan{\TilM}{1}(\cdot)$ implement non-subtractive uniform dithered quantization \cite{Gray:93}. Unlike subtractive dithered quantization, considered in, e.g., \cite{Zamir:92}, non-subtractive quantizers do not require the realization of the dithered signal to be subtracted from the quantizer output in the digital domain, resulting in a practical structure \cite{Gray:93}.   An illustration   is depicted in Figure \ref{fig:DithQuan}. 

To formulate the input-output relationship of the serial \ac{adc}, let $\DynRange$ denote the dynamic range of the quantizer, and define $\DynInt \triangleq \frac{2\DynRange}{\TilM}$ as the quantization spacing. 
The uniform quantizer is designed to operate within the dynamic range, namely,  the amplitude of the input is not larger than $\DynRange$ with sufficiently large probability. To guarantee this, we fix $\DynRange$ to be some multiple $\myEta$ of the maximal standard deviation of the input. By Chebyshev's inequality \cite[Pg. 64]{Cover:06}, for $\myEta \ge 3$  the amplitude of the input is smaller than the dynamic range with probability over $89 \%$ for any input distribution. We assume that $\myEta < \sqrt{3} \TilM$, such that the variable $\MyKappa \triangleq \myEta^2 \big(1 - \frac{ \myEta^2 }{3\TilM^2}\big)^{-1}$ is strictly positive. Note that $\eta = 3$ satisfies this requirement for any $\TilM \ge 2$, i.e., the \ac{adc} is implemented using scalar quantizers with at least one bit. 
The output of the serial scalar \ac{adc} with input sequence $y_1, y_2, \ldots, y_\lenZ$ can be written as $\Quan{\TilM}{1}\left( y_i\right)  =  q_{\lenZ}\left(y_i + z_i \right) $, where $z_1, z_2, \ldots, z_\lenZ$ are i.i.d. \acp{rv} uniformly distributed over $\left[-\frac{\DynInt}{2},\frac{\DynInt}{2} \right]$, mutually independent of the input, representing the dither signal. The function $q_{\lenZ}(\cdot)$, which implements uniform quantization, is given by 
\label{txt:UniformQnt}
\color{NewColor}
\ifsingle
\begin{equation*}
	 q_{\lenZ}(y) = \begin{cases}
-\DynRange + \DynInt\left(l +\frac{1}{2} \right)   & 
y - l \cdot \DynInt + \DynRange \in  \left[0,\DynInt \right], \quad l \in \{0,1, \ldots, \TilM - 1 \}  \\
{\rm sign}\left( y\right) \left( \DynRange - \frac{\DynInt}{2}\right)    & |y| > \DynRange.
\end{cases}
\end{equation*}  
\else
\begin{equation*}
q_{\lenZ}(y) = \begin{cases}
-\DynRange + \DynInt\left(l + \frac{1}{2} \right)   & \begin{array}{c}
  y - l \cdot \DynInt + \DynRange \in  \left[0,\DynInt \right] \\ l \in \{0,1, \ldots, \TilM - 1 \}  \end{array} \\
{\rm sign}\left( y\right) \left( \DynRange - \frac{\DynInt}{2}\right)    & |y| > \DynRange.
\end{cases}
\end{equation*}  
\fi 
\color{black}
Note that when $\TilM = 2$, the resulting quantizer is a standard one-bit sign quantizer of the form $q_{\lenZ}(y) = c \cdot {\rm sign}(y)$, where the constant $c >0$ is determined by the dynamic range $\DynRange$.

Dithered quantizers significantly facilitate the analysis, due to the  following favorable properties: 
\textcolor{NewColor}{
When the input is inside the dynamic range of the quantizer, the output can be written as the sum of the input and an additive zero-mean white quantization noise signal, which is uncorrelated with the input. This model allows us to accurately characterize the quantization system which minimizes the \ac{mse} in Theorem \ref{thm:OptimalDes}. The favorable properties of dithered quantization are also satisfied in uniform quantization  {\em without dithering} for inputs with bandlimited characteristic function, and are approximately satisfied for various families of input distributions, including the Gaussian distribution \cite{Widrow:96}. Furthermore, for those input distributions under which the favorable additive uncorrelated quantization noise model holds without dithering, the energy of the quantization noise is smaller without dithering than it is with dithering.  Consequently, while in the following analysis we assume dithered quantization, exploiting the fact that the resulting quantization noise is white and uncorrelated with the input, the proposed system can also be applied without dithering. Under input distributions for which the uncorrelated quantization noise model approximately holds, we expect lower distortion values to be achievable compared to the dithered case. This behavior is demonstrated in the simulations study in Section \ref{sec:Simulations}, where we show that applying the proposed system without dithering yields improved performance, due to the reduced energy of the quantization noise.}

\textcolor{NewColor}{
Since the dithered quantization operation can be modeled as adding uncorrelated noise, designing hardware-limited task-based quantization systems bears some similarity to linear transceiver design in \ac{mimo} communications. However, there are several fundamental differences between the two models. In hardware-limited quantization the additive quantization noise depends on the number of scalar quantizers due to the constraint on the overall number of bits, while in \ac{mimo} communications the channel noise does not depend on how many antennas are used. Furthermore, this additive noise is uncorrelated with the input but not independent, as commonly assumed in \ac{mimo} communications, and this model holds only when the quantized input is within the dynamic range. These differences between \ac{mimo} communications and hardware-limited task-based quantization result in a different system design.}

\vspace{-0.2cm}
\subsection{Hardware-Limited Task-Based Quantizer Design}
\label{subsec:OptimalHL}
\vspace{-0.1cm} 
We now characterize the hardware-limited task-based quantizer which minimizes the \ac{mse} under the system model detailed in the previous subsection. 
\textcolor{NewColor2}{Recall that here, the linear operators $\Hanalog(\cdot)$ and $\Hdigital(\cdot)$, denoting the analog combining and digital processing in Figure \ref{fig:HardLimIllust}, respectively, are represented using the matrices $\myA$ and $\myB$, respectively.}
Our characterization yields the analog combining matrix and digital processing matrix, denoted $\myA\opt$ and $\myB\opt$, respectively, and the corresponding dynamic range $\DynRange$.
Since for any quantized representation $\hat{\myTh}$, it follows from the orthogonality principle that the \ac{mse}, $\E\{\|\myTh - \hat{\myTh} \|^2 \}$, equals  the sum of the estimation error of the \ac{mmse} estimate, $\E\{\|\myTh - \myThTil \|^2 \}$, and the distortion with respect to the \ac{mmse} estimate, $\E\{\|\myThTil - \hat{\myTh} \|^2 \}$, in the following we characterize the performance of the proposed systems via the distortion with respect to $\myThTil$. 

Let $\LmmseMat$ be the \ac{mse} optimal transformation of $\myX$, namely, $\myThTil = \LmmseMat \myX$,  and let $\CovMat{\myX}$ be the covariance matrix of $\myX$, assumed to be non-singular.
Before we study the overall hardware-limited task-based quantization system, we first derive the  digital processing matrix \textcolor{NewColor}{which minimizes the \ac{mse}} for a given analog combining matrix $\myA$ and the resulting \ac{mse}, as stated in the following lemma:
\begin{lemma}
	\label{lem:ThmProof1}
\ifsingle
 	For any analog combining matrix $\myA$ and dynamic range $\DynRange$ such that $\Pr \left( \big|\left( \myA \myX\right)_l + z_l\big| > \DynRange \right) = 0$, namely, the quantizers operate within their dynamic range with probability one,   
 	the  digital processing matrix which minimizes the \ac{mse} is given by 
	\begin{equation*}
	\myB\opt \left( \myA\right)  = \LmmseMat\CovMat{\myX}\myA^T\bigg( \myA\CovMat{\myX}\myA^T + \frac{{2{\DynRange^2}}}{{3\TilM^2}}{\myI_\lenZ} \bigg)^{ - 1},
	\end{equation*}
	and the minimal achievable \ac{mse} is 
	\begin{align*}
	{\rm MSE}\left(\myA\right) 
	&= \mathop {\min }\limits_{\myB} \E\left\{ \left\| \myThTil-\hat{\myTh} \right\|^2 \right\}  \notag \\
	&\!=\! 
	{\rm Tr} \Bigg( \LmmseMat \CovMat{\myX}\LmmseMat^T\! -\! \LmmseMat\CovMat{\myX}\myA^T\!\bigg( \myA\CovMat{\myX}\myA^T\! + \!\frac{{2{\DynRange^2}}}{{3\TilM^2}}{{\bf{I}}_\lenZ} \bigg)^{ - 1}\!\!\!\!\!\myA\CovMat{\myX}\LmmseMat^T \Bigg).
	\end{align*} 	
\else
 	For any analog combining matrix $\myA$ and dynamic range $\DynRange$ such that $\Pr \left( \big|\left( \myA \myX\right)_l + z_l\big| > \DynRange \right) = 0$, namely, the quantizers operate within their dynamic range with probability one,   
 	the  digital processing matrix which minimizes the \ac{mse} is given by 
	\begin{equation*}
	\myB\opt \left( \myA\right)  = \LmmseMat\CovMat{\myX}\myA^T\bigg( \myA\CovMat{\myX}\myA^T + \frac{{2{\DynRange^2}}}{{3\TilM^2}}{\myI_\lenZ} \bigg)^{ - 1},
	\end{equation*}
	and the minimal achievable \ac{mse} is  
	\begin{align*}
	&{\rm MSE}\left(\myA\right) 
	= \mathop {\min }\limits_{\myB} \E\left\{ \left\| \myThTil-\hat{\myTh} \right\|^2 \right\}  \notag \\
	&\!=\! 
	{\rm Tr} \Bigg( \LmmseMat \CovMat{\myX}\LmmseMat^T\! -\! \LmmseMat\CovMat{\myX}\myA^T\!\bigg( \myA\CovMat{\myX}\myA^T\! + \!\frac{{2{\DynRange^2}}}{{3\TilM^2}}{{\bf{I}}_\lenZ} \bigg)^{ - 1}\!\!\!\!\!\myA\CovMat{\myX}\LmmseMat^T \Bigg).
	\end{align*} 
\fi 
\end{lemma}

{\em Proof:}
See Appendix \ref{app:ThmProof1}.

\smallskip 
\label{txt:UncorDis}
The digital processing matrix in Lemma~\ref{lem:ThmProof1} is the linear  \ac{mmse} estimator of $\myTh$ from the vector $\myA\myX + \myVec{e}$, where $\myVec{e}$ represents the quantization noise, which is white and uncorrelated with $\myA\myX$. This stochastic representation is a result of the usage of dithered quantizers with the amplitude of the input to the quantizers being within the dynamic range. The resulting requirement, namely, to set   the dynamic range $\DynRange$ such that $\Pr \left( \big|\left( \myA \myX\right)_l + z_l\big| > \DynRange \right) = 0$, is quite restrictive, as it holds only when the entires of $\myX$ have a finite support. This model can be satisfied along with the requirement for the \ac{mmse} estimator to be linear in some setups, for example, when $\myTh = \LmmseMat\myX + {\bf w}$ where ${\bf w}$ is a zero-mean random vector independent of $\myX$. However, in many scenarios of interest the entires of $\myX$ have an infinite support. In this case, there is always some probability that the input amplitude will exceed the dynamic range, resulting in some level of correlation between $\myA\myX$ and $\myVec{e}$.		

Nonetheless, in the following we use the model on which Lemma~\ref{lem:ThmProof1} is based, namely, that the output of the dithered quantizer can be written as its input corrupted by additive uncorrelated white noise,  to design hardware-limited task-based quantizers also when the entires of $\myX$ have  infinite support. 
Specifically, when the dynamic range is set such that the probability of overloading the quantizer is sufficiently small, namely,  $\Pr \left( \big|\left( \myA \myX\right)_l + z_l\big| > \DynRange \right) \approx 0$ for each $l$, then modeling  $\myA\myX$ and $\myVec{e}$ as uncorrelated becomes a reliable approximation. Therefore, in order to use Lemma \ref{lem:ThmProof1} to design hardware-limited task-based quantizers,  we explicitly require the uniform quantizers to operate within their dynamic range with high probability, as explained in Subsection \ref{subsec:SysModel}, and set the value of $\DynRange$ accordingly. In Section \ref{sec:Simulations} we numerically demonstrate that this model leads to hardware-limited task-based quantizers which are capable of approaching the performance achievable using vector quantizers for inputs with infinite support. 

We now use Lemma \ref{lem:ThmProof1} to obtain the analog combining matrix $\myA\opt$ \textcolor{NewColor}{which minimizes \ac{mse}} and the resulting system.
 Define the matrix $\tilde{\LmmseMat} \triangleq \LmmseMat \CovMat{\myX}^{1/2}$, and let $\{ \eig{\tilde{\LmmseMat},i} \}$ be its singular values arranged in a descending order. Note that for $i > {\rm rank} \big( \tilde{\LmmseMat}\big)$, $\eig{\tilde{\LmmseMat},i}  = 0$. 
 
\textcolor{NewColor}{ The hardware-limited task-based quantization system based on the structure detailed in Subsection \ref{subsec:ProblemFormulation} which minimizes the \ac{mse} under the model assumptions of  Subsection \ref{subsec:SysModel}} is given in the following theorem:
\begin{theorem}
	\label{thm:OptimalDes}
	\begin{subequations}
	\label{eqn:OptimalDes}
	For the hardware-limited quantization system based on the model detailed in Subsection \ref{subsec:SysModel}, the analog combining matrix $\myA\opt$ is given by $\myA\opt = \myMat{U}_{\myA} \myMat{\Lambda}_{\myA} \myMat{V}_{\myA}^T \CovMat{\myX}^{-1/2}$, where
	\begin{itemize}
		\item $\myMat{V}_{\myA} \in \mySet{R}^{\lenX \times \lenX}$ is the right singular vectors matrix of  $\tilde{\LmmseMat}$.
		\item  $\myMat{\Lambda}_{\myA} \in \mySet{R}^{\lenZ \times \lenX}$ is a diagonal matrix with diagonal entries  
		\begin{equation}
		\label{eqn:OptimalDesA}
		\left( \myMat{\Lambda}_{\myA}\right)_{i,i}^2 = \frac{{2{\kappa_\lenZ }}}{{3\TilM^2} \cdot \lenZ}\left( {\Wlevel  \cdot\eig{\tilde{\LmmseMat},i} - 1} \right)^ +,
		\end{equation}
		 where  $\Wlevel$ is set such that $\frac{{2{\kappa_\lenZ }}}{{3\TilM^2} \cdot \lenZ}\sum\limits_{i=1}^{\lenZ} \left( {\Wlevel  \cdot\eig{\tilde{\LmmseMat},i} - 1} \right)^ + = 1$. 
		\item $\myMat{U}_{\myA} \in \mySet{R}^{\lenZ \times \lenZ}$ is a unitary matrix which guarantees that  $\myMat{U}_{\myA}\myMat{\Lambda}_{\myA}\myMat{\Lambda}_{\myA}^T\myMat{U}_{\myA}^T$ has identical diagonal entries, namely, $\myMat{U}_{\myA}\myMat{\Lambda}_{\myA}\myMat{\Lambda}_{\myA}^T\myMat{U}_{\myA}^T$ is weakly majorized by all possible rotations of  $\myMat{\Lambda}_{\myA}\myMat{\Lambda}_{\myA}^T$ \cite[Cor. 2.1]{Palomar:07}. The matrix $\myMat{U}_{\myA}$ can be  obtained\footnote{The existence of the unitary matrix $\myMat{U}_{\myA}$ is guaranteed by \cite[Cor. 2.1]{Palomar:07}. However, this matrix is not unique as, e.g., both $\myMat{U}_{\myA}$ and $-\myMat{U}_{\myA}$ result in a rotation of $\myMat{\Lambda}_{\myA}\myMat{\Lambda}_{\myA}^T$ having identical diagonal entries. } via \cite[Alg. 2.2]{Palomar:07}. 
	\end{itemize}
	The dynamic range of the \ac{adc} is given by 
	\begin{equation}
	\label{eqn:OptimalDesGamma}
	\DynRange^2   =   \frac{ \MyKappa}{\lenZ} = \frac{ \myEta^2 }{\lenZ}\Big(1 - \frac{ \myEta^2 }{3\TilM^2}\Big)^{-1},
	\end{equation}
	and the digital processing matrix is equal to
	\begin{equation}
	\label{eqn:OptimalDesB}
	\myB\opt\left( \myA\opt\right) = \tilde{\LmmseMat} \myMat{V}_{\myA}\myMat{\Lambda}_{\myA}^T\left( \myMat{\Lambda}_{\myA} \myMat{\Lambda}_{\myA}^T+ \frac{{2{\DynRange^2}}}{{3\TilM^2}}\myI_{\lenZ} \right)^{ - 1}\!\! \myMat{U}_{\myA}^T.
	\end{equation}
	The resulting minimal achievable distortion is
	\begin{equation}
	\label{eqn:OptimalDesMSE}	
	 \E \left\{\left\|\myThTil \!-\! \hat{\myTh} \right\|^2  \right\}\! = \!
	 \begin{cases} 
	 \sum\limits_{i=1}^{\lenS}   \frac{ \eig{\tilde{\LmmseMat},i}^2} {\left(\Wlevel \cdot\eig{\tilde{\LmmseMat},i} -  1 \right)^+ \!+ \!1}, &\lenZ \!\ge\! \lenS \\
	 \sum\limits_{i=1}^{\lenZ}   \frac{ \eig{\tilde{\LmmseMat},i}^2} {\left(\Wlevel \cdot\eig{\tilde{\LmmseMat},i} -  1 \right)^+ \!+\! 1} \!+\! \sum\limits_{i\!=\!\lenZ\!+\!1}^{\lenS}  \eig{\tilde{\LmmseMat},i}^2, & \lenZ\! <\! \lenS.
	 \end{cases}
%
	\end{equation}
	\end{subequations}
\end{theorem}

{\em Proof:}
See Appendix \ref{app:ProofThmDes}.

\smallskip
\textcolor{NewColor}{
We note that, unlike task-based vector quantizers, for hardware-limited systems detailed in Subsection \ref{subsec:ProblemFormulation}, recovering the \ac{mmse} estimate $\myThTil$ in the analog domain is sub-optimal. }
Since the quantization is carried out using a serial scalar \ac{adc}, the proposed analog combining rotates the input to the \ac{adc} such that each entry has {\em identical variance}, accounting for the fact that the same quantization rule is applied to each entry. Furthermore, the analog combiner includes a waterfilling-type expression over its singular values, which accounts for the finite dynamic range of the \ac{adc}. 
In particular, the waterfilling allows the resulting system to balance the estimation and quantization errors. To see this, we note from Appendix \ref{app:ProofThmDes} that the matrix $\myMat{\Lambda}_{\myA}$ determines the dynamic range $\DynRange$. Consequently, by potentially nulling the diagonal entries corresponding to the less dominant singular values  $\{ \eig{\tilde{\LmmseMat},i} \}$, the quantization system reduces the dynamic range. This yields  more precise quantization and reduces the quantization error, at the cost of a small estimation error.

\textcolor{NewColor2}{The quantization system in Theorem \ref{thm:OptimalDes} minimizes the \ac{mse} under the model detailed in Subsection \ref{subsec:SysModel}. This model is restricted to uniform quantization mappings, as these quantizers faithfully represent typical serial \acp{adc}. The performance can be further improved by allowing for non-uniform quantizers, as shown in \cite{Shlezinger:19}, which used data-driven machine learning methods to optimize the overall system. The model-based analysis of task-based quantization with non-uniform quantizers is left for future investigation.}

Theorem \ref{thm:OptimalDes} also provides guidelines to selecting the dimensions of the output of the analog combiner, as stated in the following corollary:
\begin{corollary}
	\label{cor:SelP}
	In order to minimize the \ac{mse}, $\lenZ$ must not be larger than the rank of the covariance matrix of $\myThTil$.
\end{corollary}

{\em Proof:}
See Appendix \ref{app:ProofSelP}.

\smallskip 
Corollary \ref{cor:SelP} indicates that analog combining should project the observed vector such that the signal which undergoes the serial scalar quantization has  reduced dimensionality, not larger than the rank of the covariance of $\myThTil$. This follows since, by reducing the dimensionality of the input to the \ac{adc} while keeping the overall number of quantization levels $M$ fixed, the quantization error induced by the scalar quantization is reduced.  
The exact optimal value of $\lenZ$ is determined by the values of the non-zero  singular values  $\{ \eig{\tilde{\LmmseMat},i} \}$. In particular, the \ac{mse} expression in Theorem \ref{thm:OptimalDes} implies that decreasing $\lenZ$ below the number of non-zero singular values results in a tradeoff between improving quantization precision and increasing the estimation error.  
In the numerical analysis in Section \ref{sec:Simulations} we demonstrate that using the proposed hardware-limited task-based system, the quantization error is made negligible for relatively small $M$, and the performance approaches that of the \ac{mmse} estimator.

Finally, we show that when the quantization resolution is sufficiently large, the proposed system produces the \ac{mmse} estimate $\myThTil$. To that aim, we assume that the covariance matrix of $\myThTil$ is non-singular, thus we set $\lenZ = \lenS$. When the quantization resolution is such that $\TilM[\lenS]$ is sufficiently large, the quantization noise introduced by the \ac{adc} becomes negligible, and the output of the system can be written as
\begin{align}
\hat{\myTh} 
&\approx \myB\opt \myA\opt \myX \notag \\
&\approx \tilde{\LmmseMat} \myMat{V}_{\myA}\myMat{\Lambda}_{\myA}^T\left( \myMat{\Lambda}_{\myA} \myMat{\Lambda}_{\myA}^T  \right)^{ - 1}\myMat{\Lambda}_{\myA}\myMat{V}_{\myA}^T \CovMat{\myX}^{-1/2}\myX.
\label{eqn:HighResApprx}
\end{align}
Furthermore, for large $\TilM[\lenS]$,   the parameter $\Wlevel$ becomes $\Wlevel \approx    {   {3\TilM[\lenS]^2} \cdot \lenS} / \big({{2{\kappa_\lenS }}\sum\limits_{i=1}^{\lenS} \eig{\tilde{\LmmseMat},i} }\big)$. Thus, the diagonal entries in \eqref{eqn:OptimalDesA} become 
$\left( \myMat{\Lambda}_{\myA}\right)_{i,i}^2 \approx   { \eig{\tilde{\LmmseMat},i}} / \big({\sum\limits_{i=1}^{\lenS} \eig{\tilde{\LmmseMat},i} } \big)$. By writing the \ac{svd} $\tilde{\LmmseMat} = \myMat{U}_{\tilde{\LmmseMat}} \myMat{\Lambda}_{\tilde{\LmmseMat}} \myMat{V}_{\tilde{\LmmseMat}}^T$ in \eqref{eqn:HighResApprx}, and recalling that $\myMat{V}_{\tilde{\LmmseMat}} = \myMat{V}_{\myA}$, we have
\begin{align*}
\hat{\myTh} &\approx \myMat{U}_{\tilde{\LmmseMat}} \myMat{\Lambda}_{\tilde{\LmmseMat}} \myMat{\Lambda}_{\myA}^T\left( \myMat{\Lambda}_{\myA} \myMat{\Lambda}_{\myA}^T  \right)^{ - 1}\myMat{\Lambda}_{\myA}\myMat{V}_{\tilde{\LmmseMat}}^T \CovMat{\myX}^{-1/2}\myX \notag \\
&\stackrel{(a)}{=}  \myMat{U}_{\tilde{\LmmseMat}} \myMat{\Lambda}_{\tilde{\LmmseMat}} \myMat{V}_{\tilde{\LmmseMat}}^T \CovMat{\myX}^{-1/2}\myX \stackrel{(a)}{=} \LmmseMat \myX = \myThTil,
\end{align*}
where $(a)$ follows since for this setting of $\myMat{\Lambda}_{\myA}$,  $\myMat{\Lambda}_{\tilde{\LmmseMat}} \myMat{\Lambda}_{\myA}^T\left( \myMat{\Lambda}_{\myA} \myMat{\Lambda}_{\myA}^T  \right)^{ - 1}\myMat{\Lambda}_{\myA} = \myMat{\Lambda}_{\tilde{\LmmseMat}}$, and $(b)$ follows since $\tilde{\LmmseMat} = \LmmseMat\CovMat{\myX}^{1/2}$. Consequently, for sufficiently large quantization resolution,  $\hat{\myTh}$ approaches the \ac{mmse} estimate $\myThTil$.

\vspace{-0.2cm}
\subsection{Suboptimal Quantization Systems}
\label{subsec:Suboptimal}
\vspace{-0.1cm}
In the previous subsection we characterized the hardware-limited task-based quantization system \textcolor{NewColor}{which minimizes the \ac{mse} under the model assumptions of Subsection \ref{subsec:SysModel}}. In the following we study two suboptimal systems of interest: a system which does not carry out any processing in the analog domain, and a system which mimics the optimal vector task-based quantizer by quantizing the \ac{mmse} estimate. Our results in the following are based on the characterization of the achievable \ac{mse} for a fixed analog combining matrix $\myA$  in Lemma \ref{lem:ThmProof1}.

We begin with the suboptimal case where processing is carried out only in the digital domain.  Here, $\lenZ = \lenX$, and the analog combiner is given by $\myA = \myI_\lenX$. 	This structure accommodates the majority of systems studied in the literature in the context of tasks performed with low precision \acp{adc},  e.g., \cite{Zeitler:12,Mo:18,Corey:17,Dabeer:10,Stien:18,Yu:16, Chi:17}. The digital processing matrix which minimizes the \ac{mse} for this case and the resulting \ac{mse} are stated in the following corollary:
\begin{corollary}
	\label{cor:DigOnly} 
	When the analog combiner is $\myA = \myI_\lenX$, the minimal achievable \ac{mse} is given by
	\begin{subequations}
		\label{eqn:DigOnly}	
		\begin{align}
		\label{eqn:DigOnlyMSE}
		\hspace{-0.3cm}
		\E \left\{\left\|\myThTil \!-\! \hat{\myTh} \right\|^2  \right\} \!=\! 
		{\rm Tr}\bigg(\tilde{\LmmseMat}^T\tilde{\LmmseMat}\bigg(\myI_{\lenX} \!+\! \frac{{3\TilM[\lenX]^2}}{{2 \MyKappa[\lenX]} \SigX}\CovMat{\myX}  \bigg)^{\!-1}  \bigg),   
		\end{align}
		and the corresponding digital matrix is 
		\begin{equation}
		\label{eqn:DigOnlyB}
		\myB\opt\left( \myI_\lenX \right)   = \LmmseMat\CovMat{\myX}\bigg( \CovMat{\myX} + \frac{{2 \MyKappa[\lenX]} \SigX}{3\TilM[\lenX]^2}  {\myI_\lenX} \bigg)^{ - 1},
		\end{equation}
		where  $\SigX \triangleq {\mathop{\max}\limits_{i=1,\ldots,\lenX}}\left( \left( \CovMat{\myX}\right) _{i,i}\right)$.
	\end{subequations}
\end{corollary}

\begin{IEEEproof}
	The corollary follows directly from Lemma \ref{lem:ThmProof1}. In particular, \eqref{eqn:DigOnlyB} is obtained from the  digital processing matrix in Lemma \ref{lem:ThmProof1} by setting $\myA = \myI_\lenX$, and \eqref{eqn:DigOnlyMSE} is obtained from the resulting \ac{mse} via the matrix inversion lemma.  
\end{IEEEproof}	

\smallskip
The resulting suboptimal system bears some similarity to the task-ignorant system discussed in Section \ref{sec:Pre_Task} in the sense that quantization is carried out independently of the task. However, the system discussed in Section \ref{sec:Pre_Task} performs joint vector quantization, while \eqref{eqn:DigOnlyMSE} is achievable with a serial \ac{adc}. As a result, the system considered here can operate only when $\log M \ge \lenX$,  otherwise the scalar quantizers are assigned zero bits, while the task-ignorant system of Section \ref{sec:Pre_Task} can operate with any positive value of $\log M$.

Next, we consider a system in which the analog combining is designed to recover the \ac{mmse} estimate $\myThTil$. Here, $\lenZ = \lenS$, and $\myA = \LmmseMat$. As noted in the discussion following Theorem \ref{thm:OptimalDes}, this approach is suboptimal when working with serial scalar \acp{adc}, unlike the case with vector quantizers discussed in Section \ref{sec:Pre_Task}.  
The digital processing matrix \textcolor{NewColor}{which minimizes the \ac{mse}} for this setup and the resulting \ac{mse} are stated in the following corollary:
\begin{corollary}
	\label{cor:AnaOnly} 
	When the analog combiner is $\myA = \LmmseMat$, the minimal achievable \ac{mse} is given by
	\begin{subequations}
		\label{eqn:AnaOnly}	
		\begin{align}
		\label{eqn:AnaOnlyMSE}
		\hspace{-0.3cm}
		\E \left\{\left\|\myThTil \!-\! \hat{\myTh} \right\|^2  \right\} \!=\! 
		{\rm Tr}\bigg(\tilde{\LmmseMat}^T\tilde{\LmmseMat}\!\bigg(\!\myI_{\lenX} \!+\! \frac{{3\TilM[\lenS]^2}}{{2 \MyKappa[\lenS]} \SigS}\tilde{\LmmseMat}^T\tilde{\LmmseMat}  \bigg)^{\!-1} \! \bigg),   
		\end{align}
		and the corresponding  digital matrix is 
		\begin{equation}
		\label{eqn:AnaOnlyB}
		\myB\opt\left( \LmmseMat \right)   = \tilde{\LmmseMat}\tilde{\LmmseMat}^T\bigg( \tilde{\LmmseMat}\tilde{\LmmseMat}^T + \frac{{2 \MyKappa[\lenS]} \SigS}{3\TilM[\lenS]^2}  {\myI_\lenS} \bigg)^{ - 1},
		\end{equation}
		where  $\SigS \triangleq {\mathop{\max}\limits_{i=1,\ldots,\lenS}}\left( \E\{(\myThTil)_i^2 \} \right)$.
	\end{subequations}
\end{corollary}

\begin{IEEEproof}
	The corollary follows directly from Lemma \ref{lem:ThmProof1} using the same arguments as in the proof of Corollary \ref{cor:DigOnly}.  
\end{IEEEproof}


The approach of quantizing the \ac{mmse} estimate is in general suboptimal. When the entries of $\myThTil$ are not linearly dependent, namely, the covariance matrix of $\myThTil$ is non-singular \cite[Ch. 8.1]{Papoulis:91}, designing the analog combiner to recover the \ac{mmse} estimate \textcolor{NewColor}{minimizes the \ac{mse}} if and only if the conditions stated in the following corollary is satisfied:

\begin{corollary}
	\label{cor:OptAnalogOnly}
	When the covariance matrix of $\myThTil$ is non-singular, quantizing the \ac{mmse} estimate \textcolor{NewColor}{minimizes the \ac{mse}} if and only if the covariance matrix of  $\myThTil$ is  $\frac{1}{\lenS} \myI_{\lenS}$. 
\end{corollary}

{\em Proof:}
See Appendix \ref{app:OptAnalogOnly}.

\smallskip 
Corollary \ref{cor:OptAnalogOnly} indicates that, except for very specific statistical relationships between   $\myX$ and   $\myTh$, quantizing the entries of the \ac{mmse} estimate vector is purely suboptimal. In the numerical study presented in Section \ref{sec:Simulations} we numerically evaluate the achievable \ac{mse} of the considered systems, and illustrate that both the system proposed in Theorem \ref{thm:OptimalDes} and the suboptimal system discussed in Corollary \ref{cor:OptAnalogOnly} are able to approach the performance of the optimal vector quantizer for large number of quantization levels $M$, and that the  system of Theorem \ref{thm:OptimalDes} outperforms the suboptimal system in Corollary \ref{cor:OptAnalogOnly} for all considered values of $M$. Additionally, we illustrate that for large $\lenS$ and relatively small $M$, a notable gap in \ac{mse} is observed between the  hardware-limited task-based quantizer of Theorem \ref{thm:OptimalDes} and the suboptimal system of Corollary \ref{cor:OptAnalogOnly}.

\vspace{-0.2cm}%
\section{Guidelines for Hardware-Limited Task-Based Quantization for Arbitrary Models}
\label{sec:Asymptotic}
\vspace{-0.1cm}
\subsection{Design Guidelines}
\label{subsec:Guidelines}
\vspace{-0.1cm}
\color{NewColor}
In the previous section we characterized the distortion of hardware-limited task-based quantizers when the digital mapping $\Hdigital(\cdot)$   is a linear function. To that aim, we designed the analog combining $\Hanalog(\cdot)$ and the digital mapping $\Hdigital(\cdot)$ such that, if the quantization error induced by the serial scalar \ac{adc} is negligible, then the resulting output approaches the \ac{mmse} estimate. In scenarios where the \ac{mmse} estimate $\myThTil$ is linear, one can design linear $\Hanalog(\cdot)$ and $\Hdigital(\cdot)$ such that the resulting quantized representation approaches $\myThTil$ as $M$ increases for any value of $\lenZ \in [\lenS, \lenX]$, which denotes the dimensions of the output of the analog linear mapping. 
In particular, it was noted that when $\lenZ$ decreases to the rank of the covariance matrix of $\myThTil$, the performance of the quantization system improves. This improvement follows as more bits can be assigned to the \ac{adc}, thus reducing the error induced by scalar quantization without modifying the overall number of bits used by the system, $\log M$. 

\label{txt:Principles}
The principles used for designing the linear analog and digital mappings $\Hanalog(\cdot)$  $\Hdigital(\cdot)$  for the relationship between $\myTh$ and $\myX$ considered in Section \ref{sec:Hardware} also suggest guidelines for designing hardware-limited task-based quantization systems for arbitrary joint distributions of $\myTh$ and $\myX$ with finite-variance entries.
\color{black}
In particular,  we  propose to set $\Hanalog(\cdot)$ and  $\Hdigital(\cdot)$ according to the following guidelines:
\begin{enumerate}
	\item The mappings  $\Hanalog(\cdot),\Hdigital(\cdot)$ are such that when $\TilM$ is  large enough, $\Hdigital(\Hanalog(\myX))$ approaches the \ac{mmse} estimate $\myThTil$.
	\item The size of the output of the analog linear mapping, $\lenZ$, is as small as possible.
\end{enumerate}
\textcolor{NewColor}{
The first guideline implies that when the quantization error induced by the \ac{adc} is sufficiently small, the output of the system approaches the \ac{mmse} estimate, thus reducing the estimation error. 
The second guideline guarantees that more bits are assigned to the serial scalar \ac{adc}, thus decreasing the quantization error. Generally speaking, these guidelines provide the ability to balance quantization and estimation errors.}

We note that in some scenarios, it may not be possible to obtain or approximate the \ac{mmse} estimate, $\myThTil$, from a linear function of $\myX$ of reduced dimensions. 
For example, when $\myTh$ is estimated from the second-order statistical moments of $\myX$, as in eigen-spectrum estimation \cite{Rodrigues:17}, subspace learning \cite{Chi:17}, \ac{doa} estimation \cite{Yu:16, Liu:17}, and source localization \cite{Corey:17}. In such cases, $\myThTil$   generally cannot be obtained from a linear function of $\myX$ of reduced dimensions. 
Nonetheless, the proposed guidelines can still be applied to design the quantization system. As an illustrative example,  in the following subsection we explicitly show how these guidelines can be used for recovering the empirical covariance of an input signal. 

\vspace{-0.2cm}
\subsection{Example: Recovery from Empirical Covariance}
\label{subsec:CovarianceEst}
\vspace{-0.1cm}
 We next demonstrate how the guidelines for designing  hardware-limited task-based quantization systems discussed in Subsection \ref{subsec:Guidelines} can be applied for recovering the empirical covariance of the input. Unlike the results presented in Section \ref{sec:Hardware}, here we will not be able to explicitly characterize the resulting distortion. However, in the numerical study carried out in Section \ref{sec:Simulations} we empirically illustrate the benefits of the proposed design, and show that it outperforms processing the observations only in the digital domain, which is the more popular approach in the literature.

In particular, consider the case where the observed vector $\myX$ consists of $\Ny$ zero-mean i.i.d. $\Ky \times 1$ vectors $\{\myXi_i\}_{i=1}^{\Ny}$, i.e., $\myX = \big[\myXi_1^T, \myXi_2^T, \ldots,\myXi_{\Ny}^T  \big]^T$ and $\lenX = \Ny \cdot \Ky$. The desired vector $\myTh$ (or its \ac{mmse} estimate $\myThTil$) can be recovered from the empirical covariance of $\{\myXi_i\}_{i=1}^{\Ny}$, namely, from 
\vspace{-0.1cm}
\begin{equation*}
{\myMat{R}}_x \triangleq \frac{1}{\Ny}\sum\limits_{i=1}^{\Ny} \myXi_i\myXi_i^T. 
\vspace{-0.1cm}
\end{equation*}
For example, when $\myTh$ is the eigenspectrum of $\myX$, the \ac{mmse} estimate $\myThTil$ is obtained from ${\myMat{R}}_x $ via \cite[Eq. (22)]{Rodrigues:17}.
At first glance, the proposed guidelines cannot be used here, as, in general, for $\lenZ < \lenX$ there exists no linear transformation  $\Hanalog:\mySet{R}^{\lenX} \mapsto \mySet{R}^{\lenZ}$ such that ${\myMat{R}}_x$ can be recovered from $\Hanalog(\myX)$. However, an approximation of ${\myMat{R}}_x$ can be obtained   via the following steps:
\begin{itemize}
	\item Divide the set $\{\myXi_i\}_{i=1}^{\Ny}$ into $\Ns$ distinct sets, each consisting of $\Ks = \frac{\Ny}{\Ns}$ vectors, namely, the $l$th set is given by $\{\myXi_i\}_{i=(l-1)\Ks +1}^{l \cdot \Ks}$, $l \in \{1,2, \ldots, \Ns\}$. 
	\item Fix the analog combining such that the input to the serial scalar \ac{adc} consists of $\Ns$ vectors $\{\myZ _l\}_{l=1}^{\Ns}$, where $\myZ_l = \sum\limits_{i=(l-1)\Ks +1}^{l \cdot \Ks} \myXi_i$. This is achieved by setting $\Hanalog(\myX) = \myMat{A}\myX$, where  the entries of  $\myMat{A} \in \mySet{R}^{\Ns \cdot \Ky \times \Ny \cdot \Ky}$  are given by
	\begin{equation*}
\hspace{-0.4cm}\left(\myMat{A}\right)_{\left( {{p_1} - 1} \right){\Ky} + {q_1},\left( {{p_2} - 1} \right){\Ky} + {q_1}} = {\delta _{{q_1},{q_2}}}\sum\limits_{l = 1}^{{\Ks}} \delta _{\left( {{p_1} - 1} \right){\Ks} + l,{p_2}},  
	\end{equation*}
	for $p_1 \in \{1,2,\ldots,\Ns \}$, 	$p_2 \in \{1,2,\ldots,\Ny \}$, 	$q_1, q_2 \in \{1,2,\ldots,\Ks \}$. 
	\item In the digital domain, we approximate ${\myMat{R}}_x$ from the quantized vectors $\bar{\myZ}_l \triangleq \big[ \Quan{\TilM}{1}\left(\left(\myZ_l \right)_1  \right), \ldots, \Quan{\TilM}{1}\left(\left(\myZ_l \right)_1  \right)\big]^T$   via
	\vspace{-0.1cm}
	\begin{equation*}
	\hat{\myMat{R}}_x \triangleq \frac{1}{\Ny}\sum\limits_{l=1}^{\Ns} \bar{\myZ}_l\bar{\myZ}_l^T. 
	\vspace{-0.1cm}
	\end{equation*}
\end{itemize}
\label{txt:Rationale1}
\textcolor{NewColor}{
	The rationale behind these steps is that, as discussed in the previous subsection, it allows to balance quantization and estimation errors. To see this, we note that when the quantization error is negligible, and the number of sets $\Ns$ is sufficiently large, $\hat{\myMat{R}}_x$ approaches the true covariance matrix by the law of large numbers. Furthermore, by decreasing the number of sets $\Ns$, fewer scalar quantizers are needed, thus the quantization error induced by the serial scalar \acp{adc} is reduced. As $\Ns$ increases to $\Ny$, $\hat{\myMat{R}}_x$ approaches the true covariance of $\bar{\myZ}_l$ (up to a constant factor). 
	The proposed guidelines thus provide the ability to trade quantization and estimation errors, which is expected to be most beneficial for small values of $M$, i.e., low quantization rates, where the quantization error becomes dominant.  We  expect to have an optimal value of $\Ns$ in the range $[1, \Ny]$ for each value of $M$. This behavior, as well as the benefits of the proposed approach for finite values of $M$, are illustrated in the empirical study in Subsection \ref{subsec:SimAsym}.
} 
 In particular, in Subsection \ref{subsec:SimAsym} it is illustrated that for the problem of eigenspectrum recovery, a quantization system designed according to the above guidelines outperforms a system which performs no analog combining prior to quantization, and that the performance gap depends on the overall quantization levels $M$ and on the number of sets $\Ns$. 

\vspace{-0.2cm}%
\section{Applications and Numerical Study}
\label{sec:Simulations}
\vspace{-0.1cm}
In this section we study the application of the hardware-limited task-based quantization systems proposed in Sections~\ref{sec:Hardware}-\ref{sec:Asymptotic}, in two scenarios involving parameter acquisition from quantized measurements: 
First, in Subsection \ref{subsec:SimFinite}, we study the achievable \ac{mse} in estimating a scalar channel with finite \ac{isi} from a fixed number of quantized measurements, as in, e.g., \cite{Zeitler:12,Dabeer:10,Stien:18}, using the hardware-limited task-based quantizer proposed in Section \ref{sec:Hardware}. 
Then, in Subsection \ref{subsec:SimAsym}, we consider the problem of estimating the eigen-spectrum from a set of i.i.d. measurements, see, e.g., \cite{Rodrigues:17}, and evaluate the achievable distortion of the quantization system design detailed in Section \ref{sec:Asymptotic}.

\vspace{-0.2cm}
\subsection{\ac{isi} Channel Estimation}
\label{subsec:SimFinite}
\vspace{-0.1cm}
We first consider the estimation of a scalar \ac{isi} channel from quantized observations, as in \cite{Dabeer:10,Zeitler:12,Stien:18}.
In this scenario, the parameter vector $\myTh\Vecdim{\lenS}$ represents the coefficients of a multipath channel  with $\lenS$ taps.
The channel is estimated from a set of $\lenX = 120$ noisy observations $\myX\Vecdim{\lenX} $, given by \cite[Eq. (1)]{Zeitler:12}
\vspace{-0.1cm}
\begin{equation}
\left(\myX\Vecdim{\lenX} \right)_i = \sum\limits_{l=1}^{\lenS} \left( \myTh\Vecdim{\lenS}\right)_l \myPS_{i-l+1} + \myW_i, \qquad i \in \{1,2,\ldots,\lenX\},
\label{eqn:ChEst1}
\vspace{-0.1cm}
\end{equation} 
where $\myPS_i$ is a deterministic known training sequence, and  $\{\myW_i\}_{i=1}^{\lenX}$ are samples from an i.i.d. zero-mean unit variance Gaussian noise process independent of $\myTh$. 
In particular, the channel $\myTh$ is modeled as a zero-mean Gaussian vector with covariance matrix  $\CovMat{\myTh\Vecdim{\lenS}}$, given by  $\big( \CovMat{\myTh\Vecdim{\lenS}}\big)_{i,j} = e^{-|i-j|}$, $i,j \in \{1,2,\ldots,\lenS\} \triangleq \lenSset$, and the training sequence is given by $\myPS_i = \cos\left(\frac{2\pi i}{\lenX} \right)$ for $i >0$ and $a_i = 0$ otherwise. Note that $\myTh\Vecdim{\lenS}$ and $\myX\Vecdim{\lenX}$ are jointly Gaussian, and thus the \ac{mmse} estimator $\myThTil$ is a linear function of $\myX$. 

In the following  
we evaluate the achievable distortion of the resulting hardware-limited task-based quantization for this setup, and compare the achievable distortion to that of the optimal task-based quantizer and of the task-ignorant quantizer discussed in Section \ref{sec:Pre_Task}.
To that aim, we consider two channels: one with $\lenS = 2$ taps and one with $\lenS = 8$ taps, and let the overall number of quantization bits be $\log M \in[2\cdot\lenS, 10 \cdot\lenS]$. As $\log M$ is strictly smaller than $\lenX$,  any quantization system which is based on applying serial scalar quantization to the observation $\myX$ without any processing in the analog domain, such as the system discussed in Corollary~\ref{cor:DigOnly}, as well as the quantization systems considered in \cite{Zeitler:12,Dabeer:10}, cannot be implemented here. 

\begin{figure}
	\centering
	\includefig{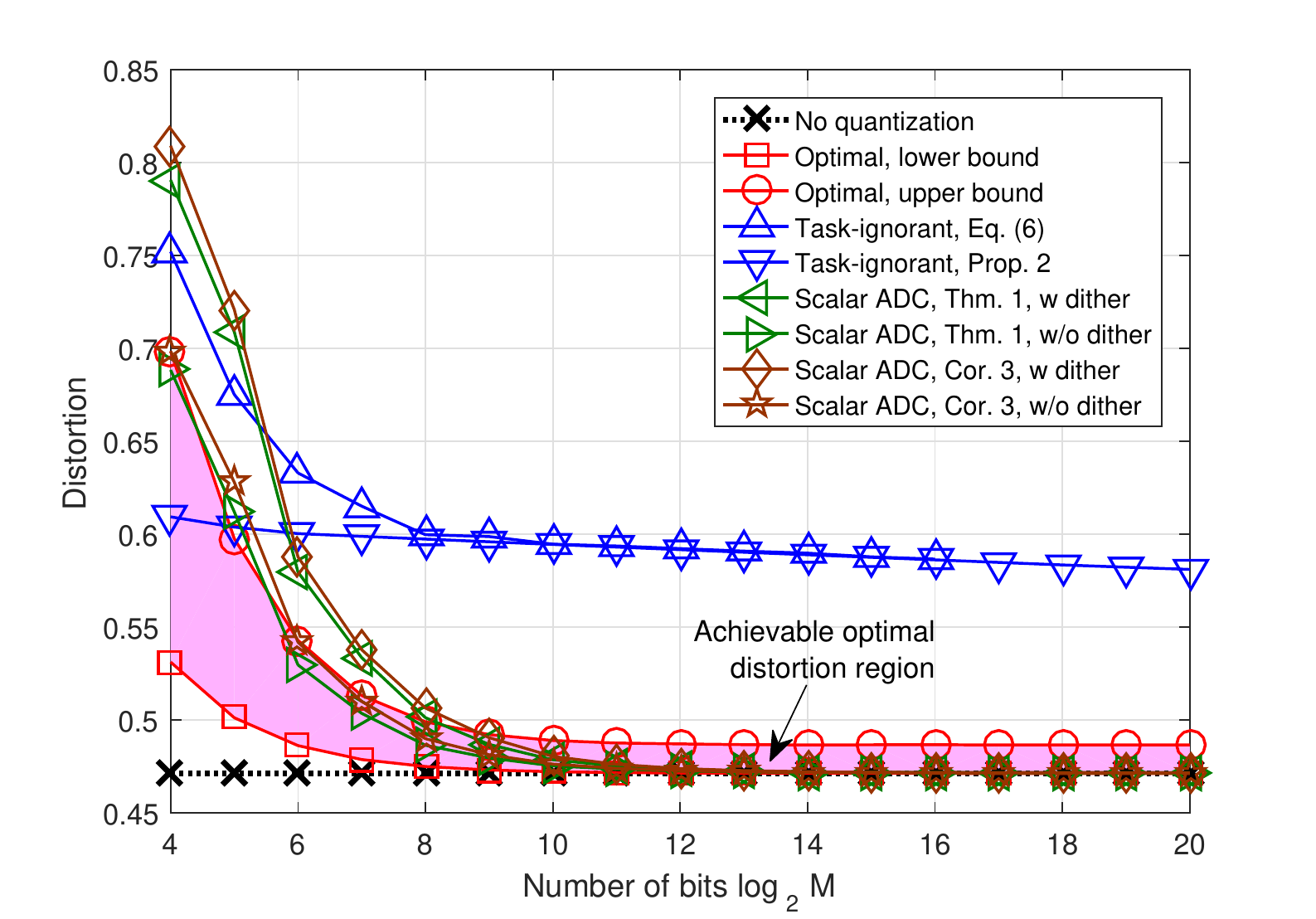}
   		\caption{Distortion comparison, channel estimation, $\lenS = 2$.}
   		\label{fig:ChEst_K2}
	\vspace{-0.4cm}
\end{figure}

\begin{figure}
	\centering
	\includefig{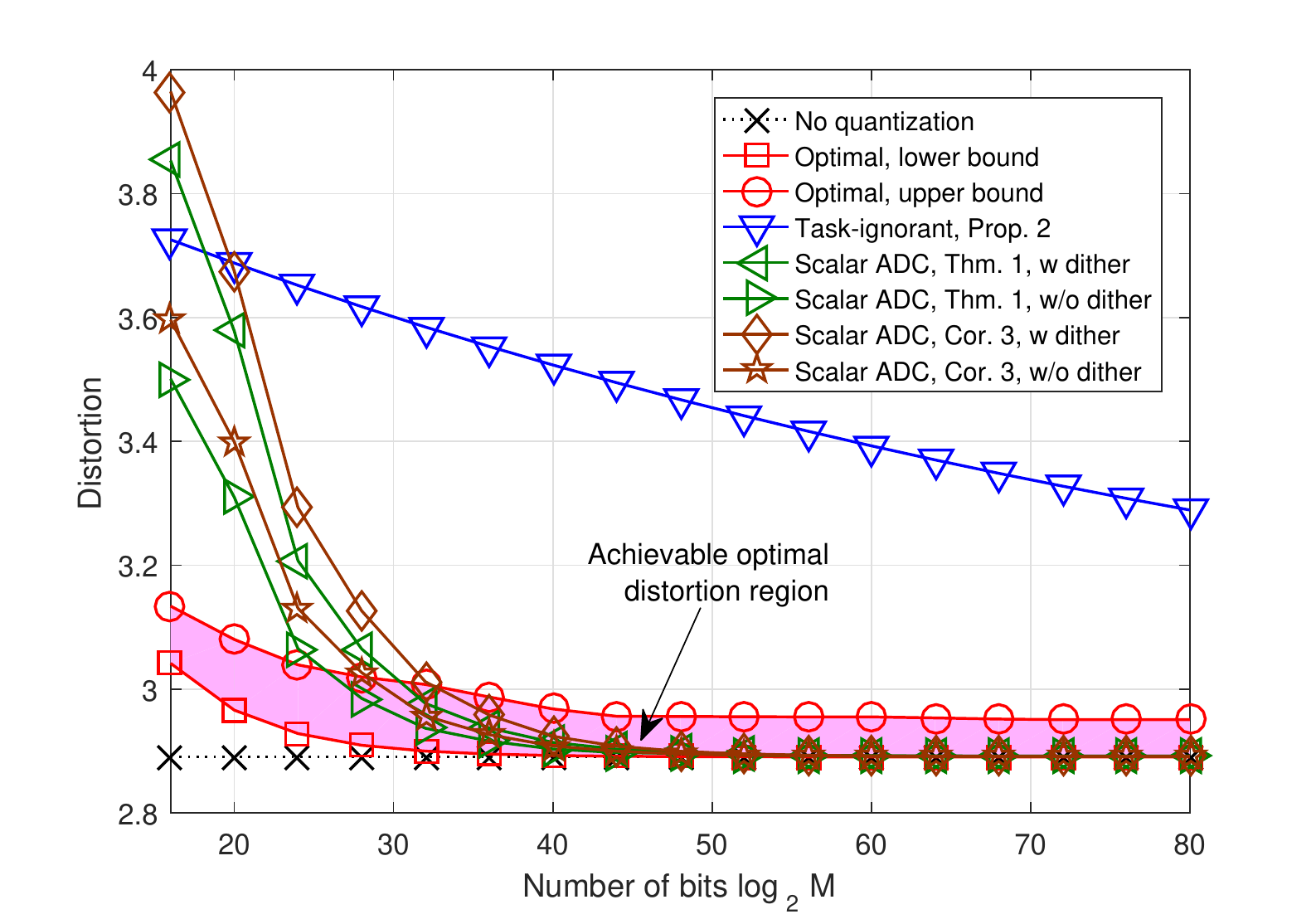}
	\caption{Distortion comparison, channel estimation, $\lenS = 8$.}
	\label{fig:ChEst_K8}
	\vspace{-0.4cm}
\end{figure}

In the numerical study we evaluate the following quantities:
\begin{itemize}
	\item The \ac{mmse} $\E\big\{ \big\| {\myTh \Vecdim{\lenS}}-{\myThTil \Vecdim{\lenS}} \big\|^2 \big\}$, which is  the optimal distortion of a system with no quantization.
	\item \label{txt:Sim2} For hardware-limited task-based quantization, we compute the achievable distortion of the  system derived in Theorem \ref{thm:OptimalDes}. Since the covariance matrix of $\myThTil$ is non-singular for the considered setup, we set $\lenZ = \lenS$ following Corollary \ref{cor:SelP}. 
	Additionally, 	we compute the  \ac{mse} of a system which recovers the \ac{mmse} estimate $\myThTil$ in the analog domain, based on Corollary \ref{cor:AnaOnly}. 
	Furthermore, since dithering increases the energy of the quantization noise, we also compute the achievable \ac{mse} of the proposed systems when the \acp{adc} implement uniform quantization without dithering. \textcolor{NewColor}{ The \ac{mse} of all these systems is computed by empirical averaging over $10000$ Monte Carlo simulations. In order to avoid cluttering, we do not depict the theoretical performance of the systems in Theorem \ref{thm:OptimalDes} and  Corollary \ref{cor:AnaOnly} with dithering, computed via \eqref{eqn:OptimalDesMSE} and \eqref{eqn:AnaOnlyMSE}, respectively. However, we note that the empirical performance depicted here coincides with the theoretical \ac{mse}.}
	\item For the {optimal task-based system}, we evaluate the bounds in Proposition \ref{pro:JointOpt1}, where the lower bound is computed using the reverse waterfilling algorithm for multivariate Gaussian sources \cite[Ch. 10.3]{Cover:06}, and the upper bound is computed via \eqref{eqn:JointOpt1},  by setting  $\Pdf{\tilde{\myC}\Vecdim{\lenS}}$  to be the optimal marginal distortion-rate distribution, which is a zero-mean Gaussian distribution whose covariance is obtained as in \cite[Ch. 10.3]{Cover:06}.
	\item For the {task-ignorant system}, we numerically evaluate the distortion in \eqref{eqn:Unconst1} by letting $\Quan{M}{\lenX}$ be the quantizer in which the codewords  are generated i.i.d. from the optimal marginal distortion-rate probability measure with respect to $\myX\Vecdim{\lenX}$, averaging the performance over $20000$ Monte Carlo simulations. 
	As this computation becomes prohibitive for large values of $M$, we evaluate \eqref{eqn:Unconst1} only for $\log M \le 16$. 
	We also compute the approximate achievable distortion of Proposition \ref{pro:TaskIgnLinear}, where the quantized output distribution is set to  the optimal marginal distortion-rate distribution for quantizing $\myX$, as proposed in the discussion following Proposition \ref{pro:TaskIgnLinear}. This quantity is computed for all considered values of $M$, and is shown to provide a good approximation of  \eqref{eqn:Unconst1} for large values of $M$.
\end{itemize}
%
%
%
 Figs.  \ref{fig:ChEst_K2}-\ref{fig:ChEst_K8} depict  the distortions for $\lenS =2$  and for $\lenS =8$, respectively. 
%
%
Observing Figs. \ref{fig:ChEst_K2}-\ref{fig:ChEst_K8}, 
we note that hardware-limited task-based quantizers substantially outperform task-ignorant vector quantization, and approach the optimal performance as $M$ increases. In particular, when each scalar quantizer uses at least five bits, i.e., $\log M \ge 5\lenS$, the quantization error becomes negligible and the overall distortion is effectively the minimum achievable estimation error, i.e., the \ac{mmse}. 

Furthermore, we note that the gap between the system of Theorem \ref{thm:OptimalDes} and the system which quantizes the \ac{mmse} estimate of Corollary \ref{cor:AnaOnly} is small for $\lenS = 2$, and becomes notable for $\lenS = 8$ at small values of $M$. However, it is emphasized that, as expected, the  system of Theorem \ref{thm:OptimalDes} outperforms the approach of quantizing the \ac{mmse} estimate, which is known to be optimal when using vector quantizers, for all considered values of $M$. 
It is also noted that the proposed hardware-limited task-based quantizers, designed assuming dithered uniform quantizers, obtain improved performance without dithering. This follows since the favorable properties of dithered quantization discussed in Subsection \ref{subsec:SysModel}, which are accounted for in the design of the systems in Section \ref{sec:Hardware}, are approximately satisfied also for non-dithered standard quantization, as noted in \cite{Widrow:96}, without the excess distortion induced by dithering.  This illustrates that our proposed design can be applied also without dithering, and that the resulting performance is improved compared to systems implementing dithered quantization. 


Regarding the achievable distortion of the non hardware-limited system discussed in Section \ref{sec:Pre_Task}, we note that the gap between the lower and the upper bounds on the optimal achievable distortion in Proposition \ref{pro:JointOpt1}, which is represented by the shaded region in the figure, narrows down from an \ac{mse} gap of $0.16$ at $\log_2 M = 4$ to $0.0158$ at $\log_2 M \ge 10$ for $\lenS = 2$, and from $0.095$ at $\log_2 M = 16$ to $0.06$ at $\log_2 M \ge 36$ for $\lenS = 8$. Since the systems proposed in Section \ref{sec:Hardware} can be obtained as a special case of the task-based vector quantizer depicted in Fig. \ref{fig:Illustration2}, combining Theorem \ref{thm:OptimalDes} and Proposition \ref{pro:JointOpt1} provides relatively tight bounds on the optimal distortion for finite quantization resolution.  We also note that the empirical distortion of the task-ignorant system is higher than the upper bound on the optimal distortion for all values of $M$, and from Fig. \ref{fig:ChEst_K2} we note  that the approximated distortion of the task-ignorant system, computed via Proposition \ref{pro:TaskIgnLinear}, provides an excellent match with the empirical distortion for  $\log_2 M > 10$, without the need to empirically average over multiple codes. 

\vspace{-0.2cm}
\subsection{Eigen-Spectrum Estimation}
\label{subsec:SimAsym}
\vspace{-0.1cm} 
We next consider the problem of estimating the eigen-spectrum of a multivariate Gaussian source from quantized measurements, as in  \cite{Rodrigues:17}. Here, the desired vector $\myTh$ represents the eigenvalues of a $\lenS \times \lenS$ covariance matrix, $\CovMat{\myX} \triangleq \myMat{U} \myMat{\Lambda} \myMat{U}^H$, where  $ \myMat{U}$ is a deterministic known unitary matrix, and $\myMat{\Lambda}$ is a diagonal matrix with diagonal entries $\left( \myMat{\Lambda}\right)_{i,i} = \left( \myTh\right)_i$, $i \in \lenSset$. 
Following \cite{Rodrigues:17}, the entries of $\myTh$ are mutually independent, and each entry $\left( \myTh\right)_i$ obeys an inverse gamma distribution with shape parameter $\alpha_i > 2$ and scale parameter  $\beta_i > 0$, $i \in \lenSset$.
The  vector $\myX$ consists of $\Ny$ random vectors $\{\myXi_i\}_{i=1}^{\Ny}$, which, given $\myTh$, are i.i.d. zero-mean Gaussian with covariance  $\CovMat{\myX}$. 

Note that for the considered scenario, $\myX$ and $\myTh$ are uncorrelated, as, by the law of total expectation,
\ifsingle
\begin{align*}
\E\{\myX \myTh^T\} = \E\{ \E\{\myX | \myTh \}  \myTh^T\} = \myMat{0}. 
\end{align*}
\else
$\E\{\myX \myTh^T\} = \E\{ \E\{\myX | \myTh \}  \myTh^T\} = \myMat{0}$.
\fi 
Hence, the linear \ac{mmse} estimator of $\myTh$ from $\myX$ is the expected value $\E\{\myTh \}$, and it thus makes little sense to design the quantization system to approach the linear \ac{mmse} estimator, which is main design principle of the systems proposed in Section \ref{sec:Hardware}. 
On the other hand, the (non-linear) \ac{mmse} estimator for this scenario is given by \cite[Thm. 4]{Rodrigues:17}
\begin{equation*}
\myThTilEnt_i \!=\! \frac{1}{\alpha_i\! - \!\frac{1}{2}\Ny \!-\! 1}\Bigg(\beta_i\! +\! \frac{1}{2}\bigg( \myMat{U}^H\bigg( \sum\limits_{l=1}^{\Ny} \myXi_l \myXi_l^T  \bigg) \myMat{U} \bigg)_{i,i}\Bigg),
\end{equation*}
$i \in \lenSset$. 
Consequently, the eigen-spectrum $\myTh$ can be estimated from the empirical covariance of $\{\myXi_i\}_{i=1}^{\Ny}$, and we thus apply the quantizer design proposed in Section \ref{sec:Asymptotic}.  
To implement the hardware-limited task-based quantizer, we divide the $\Ny$ realizations $\{\myXi_i\}_{i=1}^{\Ny}$ into $\Ns$ distinct sets of size $\Ks = \frac{\Ny}{\Ns}$. The quantization system of Section \ref{sec:Asymptotic} is then used to produce a quantized estimation of  $\frac{1}{\Ny}\sum\limits_{i=1}^{\Ny} \myXi_i\myXi_i^T$, denoted $\hat{{\myMat{R}}}_x$, using $\lenZ = \lenS \cdot \Ny$ identical scalar quantizers. The eigen-spectrum is then estimated via 
\begin{equation}
\label{eqn:EigSpecEst}
\hat{\myThTilEnt}_i = \frac{1}{\alpha_i - \frac{1}{2}\Ny - 1}\left(\beta_i + \frac{1}{2}\left( \myMat{U}^H\hat{{\myMat{R}}}_x \myMat{U} \right)_{i,i}\right),
\end{equation}
$i \in \lenSset$.  Note that when $\Ns = \Ny$, no analog combining is performed, and the quantization system results the standard approach of estimating from uniformly quantized measurements.

\begin{figure}
	\centering
	\includefig{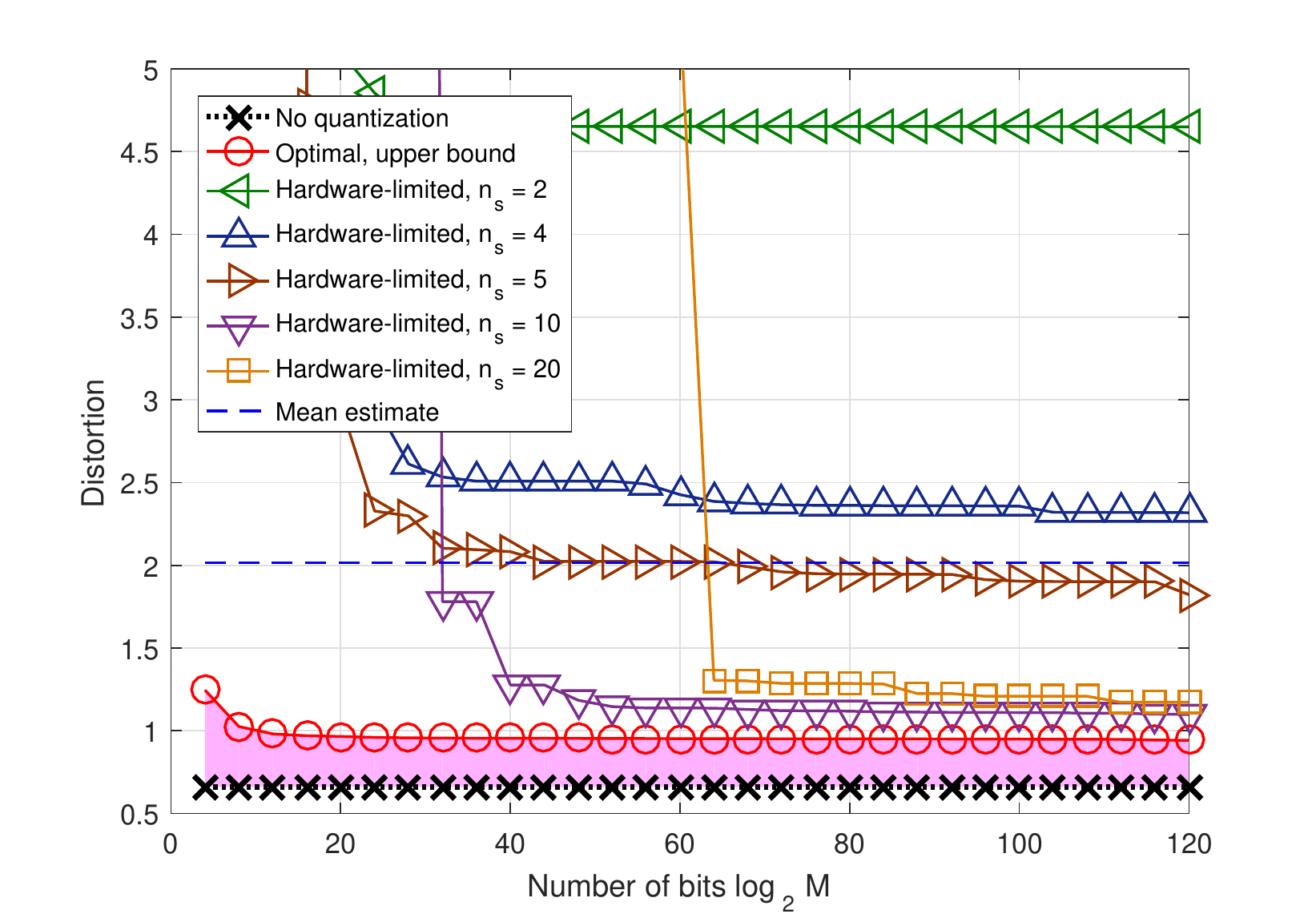}
	\caption{Distortion comparison, eigen-spectrum estimation, $\lenS = 2$.}
	\label{fig:Eigen_K2}
	\vspace{-0.4cm}
\end{figure}

\begin{figure}
	\centering
	\includefig{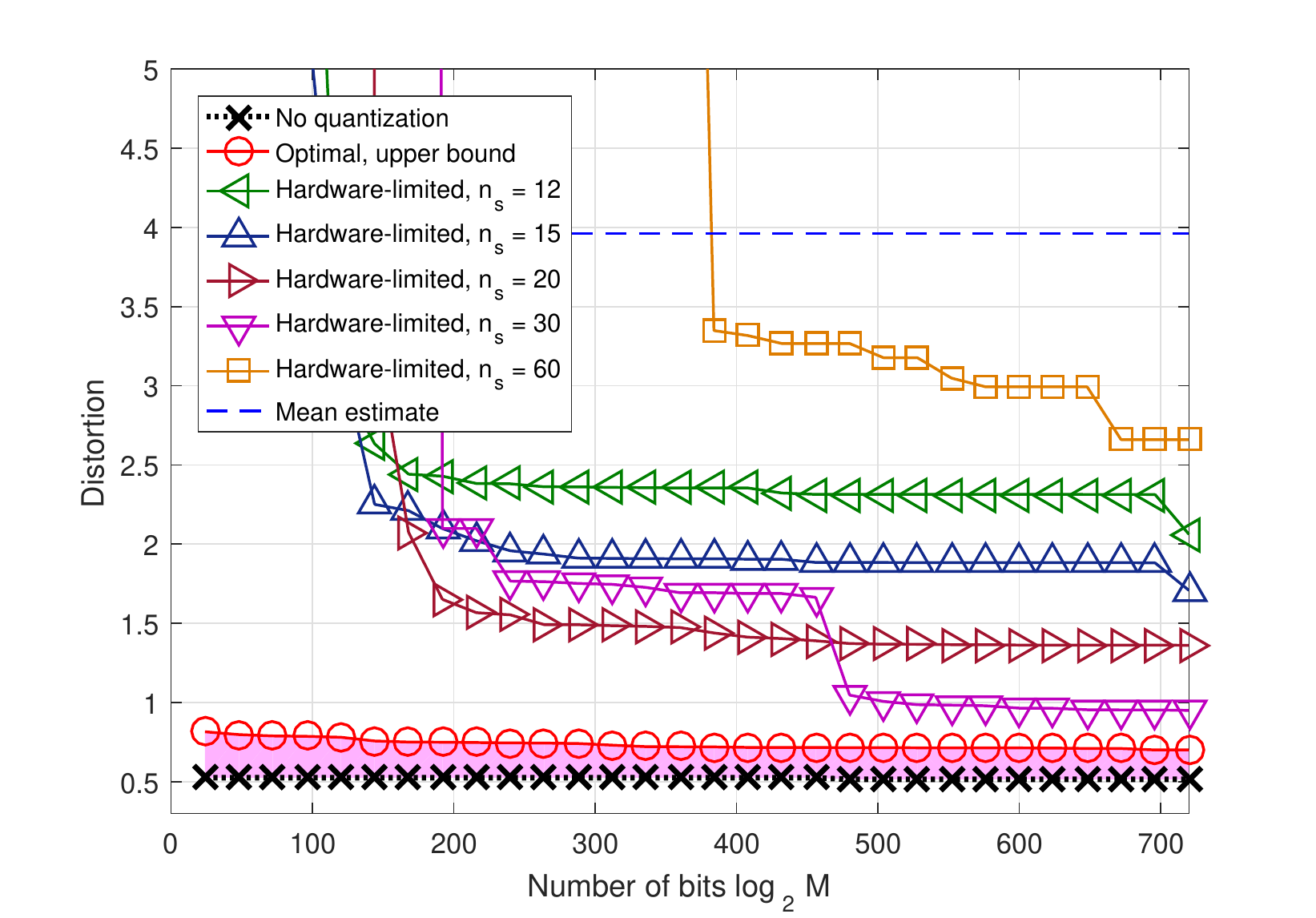}
	\caption{Distortion comparison, eigen-spectrum estimation, $\lenS = 4$.}
	\label{fig:Eigen_K4}
	\vspace{-0.4cm}
\end{figure}

In the numerical study we consider two setups: 
\begin{itemize}
	\item In the first setup, we fix $\lenS = 2$, and set the inverse gamma distribution parameters to $\{\alpha_i\}_{i=1}^{\lenS} = \{5.5,6.5\}$ and $\{\beta_i\}_{i=1}^{\lenS} = \{8.4,11.6\}$, thus, each entry of $\myTh$ has approximately unit variance \cite{Llera:16}. 
	The observed vector $\myX$ consists of $\Ny = 20$ realizations, thus $\lenX = \Ny \cdot \lenS = 40$, and the unitary matrix $\myMat{U}$  is set to the $\lenS\times\lenS$ \ac{dft} matrix.
	\item In the second setup, we fix $\lenS = 4$, and set the inverse gamma distribution parameters to $\{\alpha_i\}_{i=1}^{\lenS} = \{4,5,6,7\}$ and $\{\beta_i\}_{i=1}^{\lenS} = \{4.2,6.9,10,13.4\}$, which again results each entry of $\myTh$ having approximately unit variance. 
	The observed vector $\myX$ consists of $\Ny = 60$ realizations, thus $\lenX = 240$, and $\myMat{U}$ is set to the identity matrix.
\end{itemize} 
For both setups we numerically evaluate the achievable distortion of the quantization system proposed in Section \ref{sec:Asymptotic}, where the scalar quantization is carried out using standard (non-dithered) uniform mapping. These distortions are compared to the achievable distortion of the optimal task-based system in Proposition \ref{pro:JointOpt1}, which is computed with $\Pdf{\tilde{\myC}}$ set to the probability measure of $\myTh$. The actual minimal achievable distortion is thus upper bounded by this achievable rate, and lower bounded by the \ac{mmse} (dashed black curve in Figs. \ref{fig:Eigen_K2}-\ref{fig:Eigen_K4}). The distortions are also compared to the distortion of the optimal linear estimator, which, in this case, is the mean value $\E\{\myTh\}$ (dashed blue curve in Figs. \ref{fig:Eigen_K2}-\ref{fig:Eigen_K4}), and is thus independent of the observed signal and the quantization system. 
The distortions are computed for quantization rate $\Rate \in [0.1,3]$, resulting in $\log M \in [4, 120]$ for the first setup and $\log M \in [24, 720]$ for the second setup. 
Note that unlike the numerical study presented in Subsection \ref{subsec:SimFinite}, here we do not evaluate the performance of the task-ignorant vector quantizer, since the \ac{mmse} estimator is not linear, thus the approximation in Proposition \ref{pro:TaskIgnLinear} does not hold, and explicitly computing the distortion in \eqref{eqn:Unconst1} by Monte-Carlo simulations is not feasible for large values of $M$.

In Figure \ref{fig:Eigen_K2} we depict the achievable distortion of the proposed hardware-limited task-based quantization systems for the first scenario,  where we used $\Ns \in \{2,4,5,10,20\}$, while Figure \ref{fig:Eigen_K4} depicts the corresponding distortions for the second scenario with $\Ns \in \{12,15,20,30,60\}$.
Observing  Figs. \ref{fig:Eigen_K2}-\ref{fig:Eigen_K4}, we first note that the performance of the optimal task-based quantizer can still be approached within a small gap by the proposed hardware-limited task-based quantization by properly selecting the number of sets $\Ns$.  
In this context, we note that the best selection of $\Ns$ depends on the overall number of bits $\log M$. When $\log M$ is small, the distortion is dominated by the quantization error induced by the scalar quantizers, and thus using less partitions, which allows to assign more bits to the scalar quantizers, is beneficial. However, as $\log M$ increases, the error in  estimating the empirical covariance induced by averaging over each set becomes dominant, and thus using more sets of smaller size achieves better performance. 
In particular, for the first setup, it is observed in Figure \ref{fig:Eigen_K2} that using $\Ns = 10$ results in the smallest distortion for most considered values of $\log M$, and in fact, achieves an \ac{mse} which is less than 0.6 from the \ac{mmse} for $\log M = 40$, i.e., for quantization rate as small as $\Rate = \frac{1}{3}$. 
For the second setup, it is observed in Figure \ref{fig:Eigen_K2} that using $\Ns = 20$ is the best selection for $\log M \in [160, 480)$,  i.e., $\Rate \in \big[\frac{2}{3}, 2\big)$, while $\Ns = 30$ achieves the best performance for $\log M \ge 480$, namely $\Rate \ge 2$. 

The approach of estimating from uniformly quantized measurements without any analog combining, namely, $\Ns = \Ny$, achieves poor performance for $\log M \le 60$ when $\lenS = 2$, and for all considered values of $\log M$ when $\lenS = 4$. In fact, for the scenario in Figure \ref{fig:Eigen_K4}, the approach of estimating from uniformly quantized measurements is outperformed by the data ignorant mean estimate for most considered values of $\log M$. The results presented in this section thus demonstrate the gain of the proposed hardware-limited task-based quantization system design for systems operating with finite and relatively small number of bits $\log M$, in scenarios where the \ac{mmse} estimate is not a linear function of the observations.

%

\vspace{-0.2cm}
\section{Conclusions} 
\label{sec:Conclusions}
\vspace{-0.15cm}
In this work we studied hardware-limited task-based quantization systems, operating with practical serial scalar \acp{adc},  for finite-size signals with finite-resolution quantization. 
We characterized the  hardware-limited task-based quantizer which minimizes the \ac{mse} when the \ac{mmse} estimate of the desired signal is a linear function of the observed signal, and demonstrated that the analysis leads to design principles applicable to a much wider range of settings. We showed that, unlike when vector quantizers are used, quantizing the \ac{mmse} estimate is  generally not optimal.
Finally, we applied our results to the relevant problems of channel estimation in \ac{isi} channels and eigen-spectrum estimation. For these scenarios, we showed that the performance of the optimal task-based vector quantizer can be approached with a practical system operating with a serial scalar \ac{adc}. Furthermore, we demonstrated that   by properly accounting for the task in the design of the quantizer, hardware-limited systems can substantially outperform task-ignorant vector quantizers.
\label{txt:Concl}
The proposed hardware-limited task-based framework and the gains observed in this work for the task of signal recovery give rise to a multitude of future research directions by considering different tasks and hardware-limitations, including communications with low-resolution \acp{adc} and hardware-limited task-based quantization for classification.

%
\vspace{-0.2cm}
\begin{appendix}
%
\numberwithin{proposition}{subsection} 
\numberwithin{lemma}{subsection} 
\numberwithin{corollary}{subsection} 
\numberwithin{remark}{subsection} 
\numberwithin{equation}{subsection}	
%
%
\vspace{-0.1cm}
\subsection{Proof of Proposition \ref{pro:JointOpt1}}
\label{app:Proof1}
\vspace{-0.1cm}
The lower bound in \eqref{eqn:JointOpt1} follows from Shannon's converse \cite{Shanon:59}, which states that for any quantizer $\Quan{M}{\lenS}$, $I\left({\myThTil \Vecdim{\lenS}};{\Quan{M}{\lenS}}\left( \myThTil \right) \right)  \le \log M$, and thus
\ifsingle
\begin{align}
\mathop{\min }\limits_{\Quan{M}{\lenS}\left(  \cdot  \right)} \E\left\{ \left\| {{\myThTil \Vecdim{\lenS}}\!-\!\Quan{M}{\lenS}\left( {{\myThTil\Vecdim{\lenS}}} \right)} \right\|^2 \right\} 
&\ge \mathop{\min }\limits_{\Quan{M}{\lenS}\left(  \cdot  \right): I\left({\myTh \Vecdim{\lenS}};{\Quan{M}{\lenS}}\left( \myX\Vecdim{\lenX} \right) \right)  \le \log M} \hspace{-0.3cm}\E\left\{ \left\| {{\myThTil \Vecdim{\lenS}}\!-\!\Quan{M}{\lenS}\left( {{\myThTil\Vecdim{\lenS}}} \right)} \right\|^2 \right\} \notag \\
& 
\stackrel{(a)}{\ge}  D_{\myThTil\Vecdim{\lenS}}(\log M), 
\label{eqn:App1conv1}
\end{align}
\else
\begin{align}
&\mathop{\min }\limits_{\Quan{M}{\lenS}\left(  \cdot  \right)} \E\left\{ \left\| {{\myThTil \Vecdim{\lenS}}\!-\!\Quan{M}{\lenS}\left( {{\myThTil\Vecdim{\lenS}}} \right)} \right\|^2 \right\} 
\notag \\ 
&\qquad\ge \mathop{\min }\limits_{\Quan{M}{\lenS}\left(  \cdot  \right): I\left({\myTh \Vecdim{\lenS}};{\Quan{M}{\lenS}}\left( \myX\Vecdim{\lenX} \right) \right)  \le \log M} \hspace{-0.3cm}\E\left\{ \left\| {{\myThTil \Vecdim{\lenS}}\!-\!\Quan{M}{\lenS}\left( {{\myThTil\Vecdim{\lenS}}} \right)} \right\|^2 \right\} \notag \\
& \qquad
\stackrel{(a)}{\ge}  D_{\myThTil\Vecdim{\lenS}}(\log M), 
\label{eqn:App1conv1}
\end{align}
\fi 
where $(a)$ follows from Def. \ref{def:DistRateFunction}. 
The upper bound in \eqref{eqn:JointOpt1} is obtained by considering the quantizer in which the codewords $\{\myVec{q}_i\}_{i=1}^M$ are generated i.i.d. from the probability measure $\Pdf{\tilde{\myC}\Vecdim{\lenS}}$, and the quantizer output is set to 
\begin{equation*}
\Quan{M}{\lenS}\left( \myVec{y}\right)  = \mathop{\arg \min}\limits_{\{\myVec{q}_i\}} \|\myVec{y} -  \myVec{q}_i\|^2, \quad \forall \myVec{y} \in \mySet{R}^\lenS.
\end{equation*}
The \ac{mse} of this quantizer is given by
\ifsingle
\begin{align}
\E\left\{ \left\|\Quan{M}{\lenS}\left( {{\myThTil\Vecdim{\lenS}}} \right) \!-\! {\myThTil \Vecdim{\lenS}} \right\|^2 \right\} 
&= \E\left\{ \mathop{\min}\limits_{i} \left( \left\|\myVec{q}_i \!-\!  {\myThTil \Vecdim{\lenS}}\right\|^2\right)  \right\}  \notag \\
&= \E\left\{\E\left\{ \mathop{\min}\limits_{i} \left( \left\| \myVec{q}_i \!-\! {\myThTil \Vecdim{\lenS}}  \right\|^2\right) \Big| {\myThTil \Vecdim{\lenS}}   \right\} \right\} \notag \\
&\stackrel{(a)}{=} \E\left\{\int\limits_{0}^{\infty} \Pr\left\{ \mathop{\min}\limits_{i} \left( \left\| \myVec{q}_i \!-\! {\myThTil \Vecdim{\lenS}} \right\|^2 > t\right) \Big| {\myThTil \Vecdim{\lenS}}   \right\}dt \right\} \notag \\
&\stackrel{(b)}{=} \E \left\{ \int\limits_{0}^{\infty}  \left[ \Pr \left( \left. \left\| \tilde{\myC}\Vecdim{\lenS} \!-\! \myThTil\Vecdim{\lenS} \right\|^2 > t \right|\myThTil\Vecdim{\lenS} \right) \right]^M dt  \right\},
\label{eqn:App1achieve1}
\end{align}
\else
\begin{align}
&\E\left\{ \left\|\Quan{M}{\lenS}\left( {{\myThTil\Vecdim{\lenS}}} \right) \!-\! {\myThTil \Vecdim{\lenS}} \right\|^2 \right\} 
= \E\left\{ \mathop{\min}\limits_{i} \left( \left\|\myVec{q}_i \!-\!  {\myThTil \Vecdim{\lenS}}\right\|^2\right)  \right\}  \notag \\
&\qquad= \E\left\{\E\left\{ \mathop{\min}\limits_{i} \left( \left\| \myVec{q}_i \!-\! {\myThTil \Vecdim{\lenS}}  \right\|^2\right) \Big| {\myThTil \Vecdim{\lenS}}   \right\} \right\} \notag \\
&\qquad\stackrel{(a)}{=} \E\left\{\int\limits_{0}^{\infty} \Pr\left\{ \mathop{\min}\limits_{i} \left( \left\| \myVec{q}_i \!-\! {\myThTil \Vecdim{\lenS}} \right\|^2 > t\right) \Big| {\myThTil \Vecdim{\lenS}}   \right\}dt \right\} \notag \\
&\qquad\stackrel{(b)}{=} \E \left\{ \int\limits_{0}^{\infty}  \left[ \Pr \left( \left. \left\| \tilde{\myC}\Vecdim{\lenS} \!-\! \myThTil\Vecdim{\lenS} \right\|^2 > t \right|\myThTil\Vecdim{\lenS} \right) \right]^M dt  \right\},
\label{eqn:App1achieve1}
\end{align}
\fi 
where $(a)$ follows from \cite[Ch. 5.3]{Papoulis:91} as $\left\| \myVec{q}_i \!-\! {\myThTil \Vecdim{\lenS}}  \right\|^2$ is non-negative for any realization of ${\myThTil \Vecdim{\lenS}}$, and $(b)$ follows from the code construction. 
Eqns. \eqref{eqn:App1conv1}-\eqref{eqn:App1achieve1} prove \eqref{eqn:JointOpt1}.
\qed

\vspace{-0.2cm}
\subsection{Proof of Lemma \ref{lem:ThmProof1}}
\label{app:ThmProof1}
\vspace{-0.1cm}
To prove the lemma, we first show that the output of the \ac{adc} can be written as the sum of its input and uncorrelated noise, and then we derive the optimal digital processing matrix.

Note that when $\Pr \left( \big|\left( \myA \myX\right)_l + z_l\big| > \DynRange \right) = 0$, it follows from \cite[Thm. 2]{Gray:93} that, \textcolor{NewColor2}{since the dither signal $z_l$ is uniformly distributed over $(-\DynInt/2, \DynInt/2]$, then} the output of the scalar \ac{adc} can be written as $\myA \myX + \Qnoise$, where the quantization noise $\Qnoise$ has uncorrelated zero-mean entries with variance $\frac{\DynInt^2}{6}$. Furthermore, by \cite[Thm. 2]{Gray:93} $\Qnoise$ satisfies $\E\{\Qnoise | \myA\myX \} = \E\{\Qnoise \} = \myVec{0}$, and thus
\begin{align*}
\E \{ \myX \Qnoise^T \} &=  \E \{ \myX \E\{\Qnoise^T | \myX \}  \} \notag \\
&\stackrel{(a)}{=}  \E \{ \myX \E\{\Qnoise^T | \myA \myX \}  \} = \myMat{0}, 
\end{align*}
i.e., $\Qnoise$ is uncorrelated with  $\myX$. 
\textcolor{NewColor}{
Here, $(a)$ follows since the quantization noise  $\Qnoise$ depends on the observed vector $\myX$ only via the input to the serial scalar quantizer $\myA \myX$, and thus the conditional distribution of $\Qnoise$ given some realization $\myX = \myX'$ is equal to the conditional distribution of $\Qnoise$ given $\myA\myX = \myA \myX'$. }
Thus, the optimal digital processing matrix, which results in $\hat{\myTh}$ being the linear \ac{mmse} estimator of $\myThTil$ from  $\myA\myX + \Qnoise$, is   
\begin{align}
\myB\opt \left(\myA \right) 
&= \E\{ \myThTil \left( \myA\myX + \Qnoise\right)^T  \} \left( \E \{\left( \myA\myX + \Qnoise\right) \left( \myA\myX + \Qnoise\right)^T  \}\right)^{-1} \notag \\
&= \E\{ \LmmseMat\myX \left( \myA\myX\right)^T  \}\left(  \E \{\left( \myA\myX \right) \left( \myA\myX \right)^T  \} + \E\{\Qnoise\Qnoise^T \} \right)^{-1} \notag \\
&\stackrel{(a)}{=} \LmmseMat\CovMat{\myX} \myA ^T\bigg( \myA \CovMat{\myX} \myA ^T + \frac{\DynInt^2}{6} \myI_{\lenZ} \bigg)^{ - 1},
\label{eqn:LemAid1}
\end{align}
where $(a)$ is a result of \cite[Thm. 2]{Gray:93}.
The \ac{mse} of the linear \ac{mmse} estimate is thus given by 
\ifsingle
\begin{align}
&{\rm MSE}\left( \myA\right) 
= {\rm Tr} \left( \LmmseMat \CovMat{\myX}\LmmseMat^T \right)  - {\rm Tr} \bigg( \LmmseMat\CovMat{\myX} \myA ^T\left( \myA \CovMat{\myX} \myA ^T\! +\! \frac{\DynInt^2}{6}\myI_{\lenZ} \right)^{ - 1}\!\!\! \myA\CovMat{\myX}\LmmseMat^T \bigg) .
\label{eqn:LemAid2}
\end{align}
\else
\begin{align}
&{\rm MSE}\left( \myA\right) 
= {\rm Tr} \left( \LmmseMat \CovMat{\myX}\LmmseMat^T \right)  \notag \\
&- {\rm Tr} \bigg( \LmmseMat\CovMat{\myX} \myA ^T\left( \myA \CovMat{\myX} \myA ^T\! +\! \frac{\DynInt^2}{6}\myI_{\lenZ} \right)^{ - 1}\!\!\! \myA\CovMat{\myX}\LmmseMat^T \bigg) .
\label{eqn:LemAid2}
\end{align}
\fi 
Plugging the quantization spacing $\DynInt = \frac{2\DynRange}{\TilM}$ into \eqref{eqn:LemAid1}-\eqref{eqn:LemAid2} proves the lemma. 
\qed

\vspace{-0.2cm}
\subsection{Proof of Theorem \ref{thm:OptimalDes}}
\label{app:ProofThmDes}
\vspace{-0.1cm}
To prove the theorem, we first derive the optimal unitary rotation for a given $\myA$. Then, we characterize the optimal analog combining and the resulting \ac{mse}. We use the fact that for a fixed matrix $\myA$, the resulting \ac{mse} is given in Lemma \ref{lem:ThmProof1}.

Recall that the dynamic threshold is set to a multiple $\myEta$ of the maximal standard deviation of the quantizer input. Therefore, 
\begin{align}
\DynRange^2 &= \myEta^2 \mathop{\max}\limits_{l=1,\ldots,\lenZ} \E \left\{ \left( \left( \myA \myX\right)_l + z_l\right) ^2 \right\} \notag \\
&\stackrel{(a)}{=} \myEta^2 \mathop{\max}\limits_{l=1,\ldots,\lenZ} \E \left\{  \left( \myA \myX\right)_l ^2 \right\} + \myEta^2 \frac{\DynRange^2}{3 \TilM^2},
\label{eqn:LemAid3}
\end{align}
where $(a)$ follows since $z_l$ is independent of $\myX$ and its variance equals $\frac{\DynInt^2}{12}
=  \frac{\DynRange^2}{3 \TilM^2}$. From \eqref{eqn:LemAid3},
\begin{align}
\DynRange^2 
&= \myEta^2\left( 1 - \frac{ \myEta^2 }{3\TilM^2}\right) ^{-1}\mathop{\max}\limits_{l=1,\ldots,\lenZ} \E \left\{ \left(  \myA \myX\right)_l ^2 \right\} \notag \\
&= \MyKappa \mathop{\max}\limits_{l=1,\ldots,\lenZ} \E \left\{  \left( \myA \myX\right)_l ^2 \right\}. 
\label{eqn:DynRange}
\end{align}
Substituting \eqref{eqn:DynRange} to the expression for ${\rm MSE}(\myA)$ in Lemma~\ref{lem:ThmProof1} yields
\ifsingle
\begin{align}
&{\rm MSE}\left( \myA\right) 
= {\rm Tr} \Bigg( \LmmseMat \CovMat{\myX}\LmmseMat^T   -   \LmmseMat\CovMat{\myX} \myA ^T \bigg( \myA \CovMat{\myX} \myA ^T\!   +\! \frac{2\MyKappa}{3 \TilM^2} \mathop{\max}\limits_{l=1,\ldots,\lenZ} \E \left\{  \left( \myA \myX\right)_l ^2 \right\} \myI_{\lenZ} \bigg)^{ - 1}  \!\!\! \myA\CovMat{\myX}\LmmseMat^T \Bigg) .
\label{eqn:LemMSE}
\end{align}
\else
\begin{align}
&{\rm MSE}\left( \myA\right) 
= {\rm Tr} \Bigg( \LmmseMat \CovMat{\myX}\LmmseMat^T   -   \LmmseMat\CovMat{\myX} \myA ^T \bigg( \myA \CovMat{\myX} \myA ^T\! \notag \\
& \qquad  +\! \frac{2\MyKappa}{3 \TilM^2} \mathop{\max}\limits_{l=1,\ldots,\lenZ} \E \left\{  \left( \myA \myX\right)_l ^2 \right\} \myI_{\lenZ} \bigg)^{ - 1}  \!\!\! \myA\CovMat{\myX}\LmmseMat^T \Bigg) .
\label{eqn:LemMSE}
\end{align}
\fi 

Using \eqref{eqn:LemMSE}, we can find for each analog combining matrix $\myA$ an optimal unitary rotation, which minimizes the \ac{mse}, as stated in the following lemma:
\begin{lemma}
	\label{lem:OptRotation}
	For every matrix $\myA \in \mySet{R}^{\lenZ \times \lenX}$ there exists a unitary matrix $\myMat{U}_{\myA} \in \mySet{R}^{\lenZ \times \lenZ}$ such that 
\ifsingle
	\begin{align}
	{\rm MSE}\left( \myA\right) &\ge  {\rm MSE}\left( \myMat{U}_{\myA} \myA\right) \notag \\
	&= {\rm Tr} \Bigg( \LmmseMat \CovMat{\myX}\LmmseMat^T   -   \LmmseMat\CovMat{\myX} \myA ^T  \bigg( \myA \CovMat{\myX} \myA ^T\!   +\! \frac{2\MyKappa}{3 \TilM^2 \cdot \lenZ} {\rm Tr}\left(  \myA \CovMat{\myX} \myA ^T\right)  \myI_{\lenZ} \bigg)^{ - 1}  \!\!\! \myA\CovMat{\myX}\LmmseMat^T \Bigg).
	\label{eqn:OptRotation}
	\end{align}
\else
	\begin{align}
	 &{\rm MSE}\left( \myA\right) \ge  {\rm MSE}\left( \myMat{U}_{\myA} \myA\right) \notag \\
	 &= {\rm Tr} \Bigg( \LmmseMat \CovMat{\myX}\LmmseMat^T   -   \LmmseMat\CovMat{\myX} \myA ^T  \bigg( \myA \CovMat{\myX} \myA ^T\!  \notag \\
	 &\qquad \qquad +\! \frac{2\MyKappa}{3 \TilM^2 \cdot \lenZ} {\rm Tr}\left(  \myA \CovMat{\myX} \myA ^T\right)  \myI_{\lenZ} \bigg)^{ - 1}  \!\!\! \myA\CovMat{\myX}\LmmseMat^T \Bigg).
	 \label{eqn:OptRotation}
	\end{align}
\fi 
\end{lemma}

\begin{IEEEproof}
	Note that for any unitary matrix $\myMat{U}_{\myA}$, it follows from \eqref{eqn:LemMSE} that
\ifsingle	
	\begin{align}
	{\rm MSE}\left( \myMat{U}_{\myA} \myA\right)  
	&={\rm Tr} \left(  \LmmseMat \CovMat{\myX}\LmmseMat^T \right) \notag \\
	&  -  {\rm Tr} \Bigg( \LmmseMat\CovMat{\myX} \myA ^T  \bigg( \myA \CovMat{\myX} \myA ^T\!  +\! \frac{2\MyKappa}{3 \TilM^2} \mathop{\max}\limits_{l=1,\ldots,\lenZ} \E \left\{  \left( \myMat{U}_{\myA} \myA \myX\right)_l ^2 \right\} \myI_{\lenZ} \bigg)^{ - 1}  \!\!\!  \myA\CovMat{\myX}\LmmseMat^T \Bigg). 
	\label{eqn:LemAid4}
	\end{align} 
\else
	\begin{align}
	&{\rm MSE}\left( \myMat{U}_{\myA} \myA\right)  
	={\rm Tr} \left(  \LmmseMat \CovMat{\myX}\LmmseMat^T \right)   -  {\rm Tr} \Bigg( \LmmseMat\CovMat{\myX} \myA ^T  \bigg( \myA \CovMat{\myX} \myA ^T\! \notag \\
	& \quad +\! \frac{2\MyKappa}{3 \TilM^2} \mathop{\max}\limits_{l=1,\ldots,\lenZ} \E \left\{  \left( \myMat{U}_{\myA} \myA \myX\right)_l ^2 \right\} \myI_{\lenZ} \bigg)^{ - 1}  \!\!\!  \myA\CovMat{\myX}\LmmseMat^T \Bigg). 
	\label{eqn:LemAid4}
	\end{align}
\fi 
	For each pair of positive semi-definite symmetric matrices $\myMat{M}_1, \myMat{M}_2$, 
\ifsingle
	note that
\fi 
	the scalar function $h(\alpha) = {\rm Tr}\big( \myMat{M}_1\left( \myMat{M}_2 + \alpha \myI \right)^{-1}\big) $ is monotonically decreasing for $\alpha > 0$. Thus, by \eqref{eqn:LemAid4},  the unitary   $\myMat{U}_{\myA}$ which minimizes the \ac{mse} is given by 
	\begin{align}
	\myMat{U}_{\myA} &= \mathop{\arg\min}\limits_{\myMat{U}}  \mathop{\max}\limits_{l=1,\ldots,\lenZ} \E \left\{  \left( \myMat{U}  \myA \myX\right)_l ^2 \right\} \notag \\
	 &= \mathop{\arg\min}\limits_{\myMat{U}}  \mathop{\max}\limits_{l=1,\ldots,\lenZ}   \left( \myMat{U}  \myA \CovMat{\myX}\myA^T \myMat{U}^T  \right)_{l,l}.
	 \label{eqn:RotProb}
	\end{align}
	From majorization theory \cite[Cor. 2.4]{Palomar:07} it follows that $ \mathop{\min}\limits_{\myMat{U}}  \mathop{\max}\limits_{l=1,\ldots,\lenZ}   \left( \myMat{U}  \myA \CovMat{\myX}\myA^T \myMat{U}^T  \right)_{l,l} = \frac{1}{\lenZ} {\rm Tr} \left( \myA \CovMat{\myX}\myA^T \right)$. Furthermore, a unitary matrix which solves \eqref{eqn:RotProb} can be obtained using the iterative algorithm in \cite[Alg. 2.2]{Palomar:07}. Plugging this into \eqref{eqn:LemAid4} proves the lemma.
\end{IEEEproof}

Finally, we characterize the matrix $\myA$ which minimizes \eqref{eqn:OptRotation}. To that aim, define $\tilde{\myA} \triangleq \myA \CovMat{\myX}^{1/2}$, and recall the notation $\tilde{\LmmseMat} = \LmmseMat\CovMat{\myX}^{1/2}$. It follows from \eqref{eqn:OptRotation} that $\tilde{\myA}$ must be set to 
\ifsingle
\begin{align}
\tilde{\myA}\opt = \mathop{\arg\max}\limits_{\tilde{\myA}} {\rm Tr}\Bigg( \tilde{\LmmseMat}\tilde{\myA}^T &\bigg( \tilde{\myA}\tilde{\myA}^T\!  +\! \frac{2\MyKappa}{3 \TilM^2 \cdot \lenZ} {\rm Tr}\left( \tilde{\myA}\tilde{\myA}^T\right)   \myI_{\lenZ} \bigg)^{-1}   \tilde{\myA}\tilde{\LmmseMat}^T \Bigg). 
\label{eqn:AnalogProblem}
\end{align}
\else
\begin{align}
\tilde{\myA}\opt = \mathop{\arg\max}\limits_{\tilde{\myA}} {\rm Tr}\Bigg( \tilde{\LmmseMat}\tilde{\myA}^T &\bigg( \tilde{\myA}\tilde{\myA}^T\!  +\! \frac{2\MyKappa}{3 \TilM^2 \cdot \lenZ} {\rm Tr}\left( \tilde{\myA}\tilde{\myA}^T\right)   \myI_{\lenZ} \bigg)^{-1} \notag \\
&  \times \tilde{\myA}\tilde{\LmmseMat}^T \Bigg). 
\label{eqn:AnalogProblem}
\end{align}
\fi 
Note that the right hand side of \eqref{eqn:AnalogProblem} is invariant to replacing $\tilde{\myA}$  with $\alpha \cdot \myMat{U} \tilde{\myA}$ for any $\alpha > 0$ and for any unitary $\myMat{U}$. Consequently, we can fix ${\rm Tr}\left( \tilde{\myA}\tilde{\myA}^T\right) = 1$, and write  $\tilde{\myA} = \myMat{\Lambda}\myMat{V}^T$, where $\myMat{\Lambda} \in \mySet{R}^{\lenZ \times \lenX}$ is a diagonal matrix whose diagonal entries are arranged in a descending order, and $\myMat{V} \in \mySet{R}^{\lenX \times \lenX}$ is unitary.  Under this setting, solving \eqref{eqn:AnalogProblem} reduces to solving  
\begin{align}
&\mathop{\arg\max}\limits_{\myMat{\Lambda}, \myMat{V}} {\rm Tr}\Bigg( \tilde{\LmmseMat}^T\tilde{\LmmseMat}\myMat{V} \myMat{\Lambda}^T \bigg( \myMat{\Lambda}\myMat{\Lambda}^T\!  +\! \frac{2\MyKappa}{3 \TilM^2 \cdot \lenZ}   \myI_{\lenZ}\bigg)^{-1}  \!\!\!\myMat{\Lambda}\myMat{V}^T \Bigg), 
\notag \\
& {\text{subject to }} {\rm Tr}\left(\myMat{\Lambda}\myMat{\Lambda}^T \right) =1. \label{eqn:AnalogProblem2}
\end{align} 
We use $\myMat{\Lambda}_{\myA},\myMat{V}_{\myA} $ to denote the optimizing matrices of \eqref{eqn:AnalogProblem2}.

Let $\tilde{\myMat{\Lambda}} \triangleq \myMat{\Lambda}^T \bigg( \myMat{\Lambda}\myMat{\Lambda}^T\!  +\! \frac{2\MyKappa}{3 \TilM^2 \cdot \lenZ}   \myI_{\lenZ}\bigg)^{-1}  \!\!\!\myMat{\Lambda}$. Clearly, $\tilde{\myMat{\Lambda}}$ is a  diagonal matrix with diagonal entries 
\ifsingle
\begin{equation}
\left( \tilde{\myMat{\Lambda}}\right)_{l,l} = \frac{\left( \myMat{\Lambda}\right)_{l,l}^2 } {\left( \myMat{\Lambda}\right)_{l,l}^2 + \frac{2\MyKappa}{3 \TilM^2 \cdot \lenZ}}, \qquad l \in \{1,2,\ldots, \lenZ\},
\label{eqn:DiagEnt}
\end{equation}
\else
$\left( \tilde{\myMat{\Lambda}}\right)_{l,l} = \frac{\left( \myMat{\Lambda}\right)_{l,l}^2 } {\left( \myMat{\Lambda}\right)_{l,l}^2 + \frac{2\MyKappa}{3 \TilM^2 \cdot \lenZ}} $, for all $ l \in\{1,2,\ldots, \lenZ\}$, 
\fi 
thus, the diagonal entires of $\tilde{\myMat{\Lambda}}$ are arranged in descending order. By \cite[Thm. II.1]{Lassare:95}, the optimal unitary matrix $\myMat{V}_{\myA}$ is the right singular vectors matrix  of $\tilde{\LmmseMat}$. 
The fact that the diagonal entries of $\tilde{\myMat{\Lambda}}$ are arranged in descending order results in the optimal $\myMat{V}_{\myA}$ being  the right singular vectors matrix of $\tilde{\LmmseMat}$ instead of a permutation of this matrix \cite[Thm. II.1]{Lassare:95}.

With this setting,  \eqref{eqn:AnalogProblem2} becomes
\vspace{-0.1cm}
\begin{align} 
&\mathop{\arg\max}\limits_{\myMat{\Lambda}} \sum\limits_{i=1}^{\min(\lenS, \lenZ)} \eig{\tilde{\LmmseMat},i}^2 \cdot \frac{\left( \myMat{\Lambda}\right)_{i,i}^2 } {\left( \myMat{\Lambda}\right)_{i,i}^2 + \frac{2\MyKappa}{3 \TilM^2 \cdot \lenZ}} 
\notag \\
& {\text{subject to }} \sum\limits_{i=1}^{\lenZ} \left( \myMat{\Lambda}\right)_{i,i}^2 =1. \label{eqn:AnalogProblem3}
\vspace{-0.1cm}
\end{align} 
To solve \eqref{eqn:AnalogProblem3}, we write $\alpha_i \triangleq  \left( \myMat{\Lambda}\right)^2_{i,i}$. With this setting, \eqref{eqn:AnalogProblem3} is concave with respect to $\{\alpha_i\}_{i=1}^{\lenZ}$. Thus, the  KKT conditions are necessary and sufficient for optimality  \cite[Ch. 5.5.3]{Boyd:04}. The resulting  $\{\alpha_i\}_{i=1}^{\lenZ}$ satisfy  
%
$\alpha_i = 0$ for $i > \lenS$ and for $i \le \lenS$, 
\vspace{-0.1cm}
\begin{align}
\alpha_i  &= \left( \sqrt{\frac{2\MyKappa}{3 \TilM^2 \cdot \lenZ \cdot \beta}}\eig{\tilde{\LmmseMat},i} -  \frac{2\MyKappa}{3 \TilM^2 \cdot \lenZ } \right)^+ \notag \\
&= \frac{2\MyKappa}{3 \TilM^2 \cdot \lenZ }\left( \sqrt{\frac{3 \TilM^2 \cdot \lenZ}{2\MyKappa \cdot \beta}}\eig{\tilde{\LmmseMat},i} -  1 \right)^+, 
\vspace{-0.1cm}
\end{align}
where $\beta$ is set such that $\sum\limits_{i=1}^{\lenZ} \alpha_i  =1$. 

By defining $\Wlevel \triangleq \Big({\frac{3 \TilM^2 \cdot \lenZ}{2\MyKappa \cdot \beta}}\Big)^{1/2}$, it follows that the diagonal entries of the optimal diagonal matrix $\myMat{\Lambda}_{\myA}$  satisfy
\vspace{-0.1cm}
\begin{equation*}
\left( \myMat{\Lambda}_{\myA}\right)_{i,i}^2 = 
\begin{cases}
\frac{2\MyKappa}{3 \TilM^2 \cdot \lenZ }\left(\Wlevel \cdot\eig{\tilde{\LmmseMat},i} -  1 \right)^+, & i \le \min(\lenS,\lenZ) \\
0 & i > \min(\lenS,\lenZ),
\end{cases}
\vspace{-0.1cm}
\end{equation*}
where $\Wlevel> 0 $ is set such that  $\frac{2\MyKappa}{3 \TilM^2 \cdot \lenZ }\sum\limits_{i=1}^{\lenZ} \big(\Wlevel \cdot \eig{\tilde{\LmmseMat},i} -  1 \big)^+ =1$. 

The optimal analog combining is thus given by $\myA\opt = \myMat{U}_{\myA} \myMat{\Lambda}_{\myA} \myMat{V}_{\myA}^T \CovMat{\myX}^{-1/2}$.
Note that under this setting, the dynamic range in \eqref{eqn:DynRange} is given by 
$\DynRange^2   =   \frac{ \MyKappa}{\lenZ} {\rm Tr}\big( \myMat{\Lambda}_{\myA}\myMat{\Lambda}_{\myA}^T\big) =  \frac{ \MyKappa}{\lenZ}$, proving  \eqref{eqn:OptimalDesGamma}.
The resulting optimal \ac{mse} can be written as
\ifsingle
	\begin{align}
	{\rm MSE}\left(\myA\opt \right)
	&= {\rm MSE}\big(\tilde{\myA}\opt \big) \notag \\
	&= {\rm Tr}\left(  \LmmseMat \CovMat{\myX}\LmmseMat^T \right) - \sum\limits_{i=1}^{\lenS} \eig{\tilde{\LmmseMat},i}^2 \cdot \frac{\left( \myMat{\Lambda}_{\myA}\right)_{i,i}^2 } {\left( \myMat{\Lambda}_{\myA}\right)_{i,i}^2 + \frac{2\MyKappa}{3 \TilM^2 \cdot \lenZ}} \notag\\
	&= \sum\limits_{i=1}^{\lenS}  \eig{\tilde{\LmmseMat},i}^2 - \sum\limits_{i=1}^{\min(\lenS,\lenZ)}\eig{\tilde{\LmmseMat},i}^2  \frac{\big(\Wlevel \cdot\eig{\tilde{\LmmseMat},i} -  1 \big)^+   } {\big(\Wlevel \cdot\eig{\tilde{\LmmseMat},i} -  1 \big)^+ + 1}. 
	\label{eqn:ProofMSE2}
	\end{align}
\else
\vspace{-0.1cm}
	\begin{align}
	&{\rm MSE}\left(\myA\opt \right)
	 = {\rm MSE}\big(\tilde{\myA}\opt \big) \notag \\
	 &= {\rm Tr}\left(  \LmmseMat \CovMat{\myX}\LmmseMat^T \right) - \sum\limits_{i=1}^{\lenS} \eig{\tilde{\LmmseMat},i}^2 \cdot \frac{\left( \myMat{\Lambda}_{\myA}\right)_{i,i}^2 } {\left( \myMat{\Lambda}_{\myA}\right)_{i,i}^2 + \frac{2\MyKappa}{3 \TilM^2 \cdot \lenZ}} \notag\\
	 &= \sum\limits_{i=1}^{\lenS}  \eig{\tilde{\LmmseMat},i}^2 - \sum\limits_{i=1}^{\min(\lenS,\lenZ)}\eig{\tilde{\LmmseMat},i}^2  \frac{\big(\Wlevel \cdot\eig{\tilde{\LmmseMat},i} -  1 \big)^+   } {\big(\Wlevel \cdot\eig{\tilde{\LmmseMat},i} -  1 \big)^+ + 1}.
	 \vspace{-0.1cm}
	 \label{eqn:ProofMSE2}
	 \end{align}
\fi 
 Since $\eig{\tilde{\LmmseMat},i}^2 \cdot \left(1-  \frac{\left(\Wlevel \cdot\eig{\tilde{\LmmseMat},i} -  1 \right)^+   } {\left(\Wlevel \cdot\eig{\tilde{\LmmseMat},i} -  1 \right)^+ + 1} \right)  = \frac{ \eig{\tilde{\LmmseMat},i}^2} {\left(\Wlevel \cdot\eig{\tilde{\LmmseMat},i} -  1 \right)^+ + 1}$, it follows that \eqref{eqn:ProofMSE2} results in
 \vspace{-0.1cm}
 \begin{align*}
 &{\rm MSE}\left(\myA\opt \right) = 
 \begin{cases} 
 \sum\limits_{i=1}^{\lenS}   \frac{ \eig{\tilde{\LmmseMat},i}^2} {\left(\Wlevel \cdot\eig{\tilde{\LmmseMat},i} -  1 \right)^+ + 1}, &\lenZ \ge \lenS \\
 \sum\limits_{i=1}^{\lenZ}   \frac{ \eig{\tilde{\LmmseMat},i}^2} {\left(\Wlevel \cdot\eig{\tilde{\LmmseMat},i} -  1 \right)^+ + 1} + \sum\limits_{i=\lenZ+1}^{\lenS}  \eig{\tilde{\LmmseMat},i}^2, & \lenZ < \lenS.
 \end{cases}  
 \vspace{-0.1cm}
  \end{align*}
Combining this with the design of   $\myA\opt$ proves the theorem.
\qed

\vspace{-0.2cm}
\subsection{Proof of Corollary \ref{cor:SelP}}
\label{app:ProofSelP}
\vspace{-0.1cm}
We note that the \ac{mse} in Theorem \ref{thm:OptimalDes} decreases as $\Wlevel$ increases. Therefore, $\lenZ$ must be set such that $\Wlevel$ is maximized. Let $r$ denote the number of non-zero singular values $\{ \eig{\tilde{\LmmseMat},i} \}$. From the definition of $\tilde{\LmmseMat}$ it follows that $r$ is also the rank of the covariance matrix of $\myThTil$. When $\lenZ \ge r$,  $\Wlevel$ is set such that 
\vspace{-0.1cm}
\begin{equation}
\frac{{2{\kappa_\lenZ }}}{{3\TilM^2} \cdot \lenZ}\sum\limits_{i=1}^{r} \left( {\Wlevel  \cdot\eig{\tilde{\LmmseMat},i} - 1} \right)^ + = 1. 
\vspace{-0.1cm}
\label{eqn:Psel1}
\end{equation}
Since $\TilM$ grows exponentially with $1/\lenZ$, it follows that as $\lenZ$ decreases, the value of $	\frac{{2{\kappa_\lenZ }}}{{3\TilM^2} \cdot \lenZ}$, also decreases, and thus $\Wlevel$ is maximized for $\lenZ \ge r$ when $\lenZ = r$,  proving the corollary. 
\qed

\vspace{-0.2cm}
\subsection{Proof of Corollary \ref{cor:OptAnalogOnly}}
\label{app:OptAnalogOnly}
\vspace{-0.1cm}
Since the optimal $\myA\opt$ for any $\lenZ$, including $\lenZ = \lenS$, is given in Theorem \ref{thm:OptimalDes}, in the following we find a necessary and sufficient condition for which $\myA\opt = \LmmseMat$, or equivalently, $\myMat{U}_{\myA}\myMat{\Lambda}_{\myA} \myMat{V}_{\myA}^T = \tilde{\LmmseMat}$. Note that  $\myMat{V}_{\myA}$ already equals the right-singular matrix of $\tilde{\LmmseMat}$, and thus $\myA = \LmmseMat$ is optimal if and only if $\myMat{U}_{\myA} \myMat{\Lambda}_{\myA}\myMat{\Lambda}_{\myA}^T\myMat{U}_{\myA} = \tilde{\LmmseMat}\tilde{\LmmseMat}^T$. 
This condition is satisfied only if 
\begin{equation}
\label{eqn:CorProof0}
\left(\myMat{\Lambda}_{\myA}\right)_{i,i}^2 = \eig{\tilde{\LmmseMat},i}^2, \quad \forall i \in \lenSset.
\end{equation} 
Since $\tilde{\LmmseMat}\tilde{\LmmseMat}^T$ is the covariance matrix of the \ac{mmse} estimate $\myThTil$, the fact that the covariance matrix is non-singular implies that  $\eig{\tilde{\LmmseMat},i} \ne 0$ for all $i \in \lenSset$. Combining this with \eqref{eqn:OptimalDesA} results in
\ifsingle
\begin{align}
\eig{\tilde{\LmmseMat},i}^2 
&= \frac{{2{\MyKappa[\lenS] }}}{{3\TilM[\lenS]^2} \cdot \lenS}\left( {\Wlevel  \cdot\eig{\tilde{\LmmseMat},i} - 1} \right)^ + \notag \\
&=\frac{{2{\MyKappa[\lenS] }}}{{3\TilM[\lenS]^2} \cdot \lenS}\left( {\Wlevel  \cdot\eig{\tilde{\LmmseMat},i} - 1} \right).
\label{eqn:CorProof1}
\end{align}
\else
\begin{align}
\hspace{-0.2cm}
\eig{\tilde{\LmmseMat},i}^2 
&\!=\! \frac{{2{\MyKappa[\lenS] }}}{{3\TilM[\lenS]^2} \cdot \lenS}\left( {\Wlevel  \cdot\eig{\tilde{\LmmseMat},i}\! - \!1} \right)^ + \!=\!\frac{{2{\MyKappa[\lenS] }}}{{3\TilM[\lenS]^2} \cdot \lenS}\left( {\Wlevel  \cdot\eig{\tilde{\LmmseMat},i} \!-\! 1} \right).
\label{eqn:CorProof1}
\end{align}
\fi 
Consequently, $ {\Wlevel  \cdot\eig{\tilde{\LmmseMat},i} - 1} > 0$ for all $i \in \lenSset$, and thus, from the condition $\frac{{2{\kappa_\lenZ }}}{{3\TilM^2} \cdot \lenZ}\sum\limits_{i=1}^{\lenZ} \big( {\Wlevel  \cdot\eig{\tilde{\LmmseMat},i} - 1} \big)^ + = 1$, we have that $\Wlevel = {\lenS}\left(\frac{{3\TilM[\lenS]^2}}{{2{\MyKappa[\lenS] }}} + 1 \right)\Big({\sum\limits_{i=1}^{\lenS}\eig{\tilde{\LmmseMat},i} }\Big)^{-1}$. Plugging this into \eqref{eqn:CorProof1} results in 
\vspace{-0.2cm}
\begin{equation}
\eig{\tilde{\LmmseMat},i}^2 = {\eig{\tilde{\LmmseMat},i}}\Big({\sum\limits_{i=1}^{\lenS}\eig{\tilde{\LmmseMat},i}}\Big)^{-1}, \quad \forall i \in \lenSset.
\label{eqn:CorProof2}
\vspace{-0.1cm}
\end{equation} 
Note that \eqref{eqn:CorProof2} is satisfied if and only if  $\eig{\tilde{\LmmseMat},i} = \frac{1}{\sqrt{\lenS}}$.  As a result,  $\myMat{\Lambda}_{\myA}$ in  \eqref{eqn:CorProof0} has identical diagonal entries,  $\myMat{U}_{\myA}$ can be any permutation matrix, and both the condition on $\myMat{U}_{\myA}$ in Theorem \ref{thm:OptimalDes} as well as the condition  $\myMat{U}_{\myA} \myMat{\Lambda}_{\myA}\myMat{\Lambda}_{\myA}^T\myMat{U}_{\myA} = \tilde{\LmmseMat}\tilde{\LmmseMat}^T$ are satisfied here. Consequently, the condition $\eig{\tilde{\LmmseMat},i} = \frac{1}{\sqrt{\lenS}}$ is not only necessary for $\myA = \LmmseMat$ to be optimal, but it is also sufficient.
Noting that in this case, the covariance matrix of $\myThTil$ is $\frac{1}{\lenS} \myI_\lenS$, concludes the proof of the corollary.
\qed

\end{appendix}

\end{document}